\documentclass[preprint,floatfix]{revtex4}%
\usepackage{graphicx}
\usepackage{amsmath}
\usepackage{amsfonts}
\usepackage{amssymb}%
\begin{document}
\title{Energy-level diagrams and their contribution to fifth-order Raman and
second-order infrared responses: Distinction between relaxation mechanisms by
two-dimensional spectroscopy}
\author{Ko Okumura}
\email{okumura@phys.ocha.ac.jp}
\affiliation{Department of Physics, Graduate School of Humanities and Sciences, Ochanomizu
University, 2--1--1, Otsuka, Bunkyo-ku, Tokyo 112-8610, Japan\linebreak%
\ and\linebreak Physique de la Mati\`{e}re Condens\'{e}e, Coll\`{e}ge de
France, 11 place Marcelin-Berthelot, 75231 Paris cedex 05, France}
\author{Yoshitaka Tanimura}
\email{tanimura@ims.ac.jp}
\affiliation{Theoretical Studies, Institute for Molecular Science, Okazaki, Aichi 444-8585, Japan}
\date{\today}

\begin{abstract}
We develop a Feynman rule for energy-level diagrams emphasizing their
connections to the double-sided Feynman diagrams and physical processes in the
Liouville space. Thereby we completely identify such diagrams and processes
contributing to the two-dimensional response function in the Brownian
oscillator model. We classify such diagrams or processes in quartet and
numerically present signals separately from each quartet of diagrams or
Liouville-space processes. We find that the signal from each quartet is
distinctly different from the others; we can identify each peaks in frequency
domain with a certain quartet. This offers the basis for analyzing and
assigning actual two-dimensional peaks and suggests the possibility of
Liouville-space-path selective spectroscopy. As an application we demonstrate
an example in which two familiar homogeneous mechanisms of relaxation are
distinguished by existence or non-existence of certain peaks on the
two-dimensional map; appearance or disappearance of certain peak is sensitive
to the coupling mechanism. We also point out some confusion in the literature
with regard to inclusion of relaxation effects.

\textbf{\bigskip\noindent}\noindent

\noindent\textbf{TITLE RUNNING\ HEAD: Energy-level diagrams and their
contribution to 2D spectroscopy}

\end{abstract}
\maketitle

\section{Introduction}

The use of ultrashort laser pulse to probe the properties of molecules has
been propelled by the rapid advances in laser measurement
techniques.\cite{Mukamel} Recently, two-dimensional (2D) vibrational
spectroscopy has been actively studied, where the spectral properties of
multi-body correlation functions of polarizability (2D Raman spectroscopy)
\cite{TM93,COT,Duppen01,KA1,GT,Miller02,Saito,Saito02,Fleming02,Reichman02,Cao02,Stratt02,Keyes}
or dipole moment (2D infrared spectroscopy) \cite{Hoch98,Hamm01,Tok01,Cho01}
are measured. The 2D technique provides information about the inter- and
intra-molecular interactions which cause energy relaxations.
\cite{Mukamel00,Fourkas01,Cho02,Hoch02,Wright}

Theoretically, optical responses of molecular vibrational motions have been
studied mainly by either an oscillator model \cite{Oxtoby} or energy level
model. \cite{Redfield} The oscillator model utilizes molecular coordinates to
describe molecular motions. This description is physically intuitive since
optical observables (dipole moments or Raman polarizability) are also
described by molecular coordinates; the effects of relaxation, which are
caused by interactions of the coordinate with some other degrees of freedom,
are rather easy to be included. As long as the potential is harmonic or nearly
harmonic, signals can be calculated analytically.
\cite{TM93,OTWANH,OT,OTCPL,CPLfreq,Suzuki01,Suzuki02}

On the contrary, the energy-level model employs the energy eigen functions of
a molecular motion but is physically equivalent to the oscillator model.
Accordingly, laser interactions are described by transitions between the
energy levels; the optical processes, including the time-ordering of laser
pulses, are conveniently described by diagrams such as Albrecht diagrams,
\cite{LeeAlbrecht} or double-sided Feynman diagrams. \cite{Mukamel} Although
the inclusion of relaxation processes from physical insight is less intuitive
and is restricted to some special cases, this model has the advantage in
identifying peak positions of optical signal in frequency domain.
\cite{Steffen,Fourkas97,Tominaga01,Tok01b} The anharmonicity of potential and
nonlinear mode-mode coupling are also easily taken into account. Phase
matching conditions, which chose a specific Liouville path contribution by the
configuration of Laser beams, \cite{Mukamel} is also easy to take into
account. In the oscillator model or molecular dynamical simulations, the phase
matching condition can be done only after calculating entire response
functions. \cite{KTCPL}

The rate of increase in the number of diagrams, however, with the increase of
laser interactions is severer in the energy-level model; this becomes serous
practical problem for multi-dimensional spectroscopy, where many laser
interactions are included. For example, more than 16 diagrams are involved in
the lowest order in the third-order anhamonicity in fifth-order Raman while
all the diagrams can be represented by a single field-theoretic diagram in the
oscillator model. \cite{OT}

In this paper we try to bridge the two complementary models by transferring
some results obtained in the oscillator-model to the energy-level language.
Although we lose the simplicity (e.g. small number of diagrams) we gain in an
insight into optical processes; we can assign each peaks in certain optical
\ or Liouville-space processes. The resulting energy-level Feynman rule for
the oscillator system allows inclusion of relaxation mechanism in an ad hoc
way. As an application, we compare two system with different damping
constants. This example reveals that existence of certain peaks in 2D
spectroscopic map sensitively depends on the relaxation model.

\section{Interaction of energy level diagrams}

We consider a molecular vibrational motion described by a single molecular
coordinate $Q$. In the energy-level representation, the Hamiltonian is
expressed as%
\begin{equation}
H_{0}=\hbar\Omega\left(  a^{\dag}a+\frac{1}{2}\right)  \label{H0}%
\end{equation}
where $a$ and $a^{\dag}$ are the creation and annihilation operators and
\begin{equation}
Q=\sqrt{\frac{\hbar}{2M\Omega}}\left(  a+a^{\dag}\right)  \label{Q}%
\end{equation}
for the system with the mass $M$. The energy level of this harmonic system is
given by $E_{n}=\hbar\Omega_{n}$ with $\Omega_{n}=(n+1/2)\Omega$ for which we
introduce the frequency difference $\Omega_{mn}=\Omega_{m}-\Omega_{n}.$ If the
system interacts with the laser field $E(t)$, it is governed by the full
Hamiltonian,%
\begin{equation}
H(t)=\left\{
\begin{array}
[c]{cc}%
H_{0}+\mu E(t) & \text{(IR)}\\
H_{0}+\alpha E(t)^{2} & \text{(Raman)}%
\end{array}
\right.  \label{Ht}%
\end{equation}
where $\mu$ is the dipole for infrared (IR)\ and $\alpha$ is the
polarizability for Raman spectroscopy. Both operators can be expanded as
\begin{equation}
x=x_{0}+x_{1}Q+\frac{1}{2!}x_{2}Q^{2}+\frac{1}{3!}x_{3}Q^{3}+\cdots,
\label{xexp}%
\end{equation}

We consider the response function, which is pertinent to the 2D second-order
IR (for non isotropic media) or the 2D fifth-order Raman spectroscopy,
\begin{align*}
&  R^{(2)}(T_{1},T_{2})\\
&  =\theta\left(  t_{3}-t_{2}\right)  \theta\left(  t_{2}-t_{1}\right)
\left\langle \left[  \left[  x\left(  t_{3}\right)  ,\frac{i}{\hbar}x\left(
t_{2}\right)  \right]  ,\frac{i}{\hbar}x\left(  t_{1}\right)  \right]
\right\rangle
\end{align*}
where $x(t)$ is the Heisenberg operator of $x$ for the non-interacting
Hamiltonian $H_{0}$ and $\left\langle O\right\rangle \equiv\mathrm{Tr}\left[
\rho_{0}O\right]  $ with $\rho_{0}=e^{-\beta H_{0}}/\mathrm{Tr}\left[
e^{-\beta H_{0}}\right]  $ (when we include the effect of dissipation at the
level of Hamiltonian, $H_{0}$ includes the bath Hamiltonian and the
system-bath interaction). The operator $x$ stands for $\mu$ (IR) or $\alpha$
(Raman). Generalization to the combined IR and Raman cases such as
$\left\langle \left[  \left[  \mu\left(  t_{3}\right)  ,\mu\left(
t_{2}\right)  \right]  ,\alpha\left(  t_{1}\right)  \right]  \right\rangle $
\cite{PC,Chorev,Wright,WrightA} will also be treated below.

$R^{(2)}(T_{1},T_{2})$ for the harmonic system can be expanded in terms of $Q$
by Eq. (\ref{xexp}). The leading order is given as%
\begin{equation}
R^{(2)}(T_{1},T_{2})=\left(  \frac{i}{\hbar}\right)  ^{2}\frac{x_{1}^{2}x_{2}%
}{2}\left(  R_{1}+R_{2}+R_{3}\right)  , \label{R2}%
\end{equation}
where%
\begin{align}
R_{1}  &  =\left\langle \left[  \left[  Q^{2}\left(  T_{1}+T_{2}\right)
,Q\left(  T_{1}\right)  \right]  ,Q\left(  0\right)  \right]  \right\rangle
\nonumber\\
R_{2}  &  =\left\langle \left[  \left[  Q\left(  T_{1}+T_{2}\right)
,Q^{2}\left(  T_{1}\right)  \right]  ,Q\left(  0\right)  \right]
\right\rangle \nonumber\\
R_{3}  &  =\left\langle \left[  \left[  Q\left(  T_{1}+T_{2}\right)  ,Q\left(
T_{1}\right)  \right]  ,Q^{2}\left(  0\right)  \right]  \right\rangle
.\nonumber
\end{align}
with%
\begin{align}
t_{3}-t_{2}  &  =T_{2}\label{td}\\
t_{2}-t_{1}  &  =T_{1},\nonumber
\end{align}

\subsection{Raman spectroscopy}

\begin{figure}
\begin{center}
\includegraphics[scale=0.7]{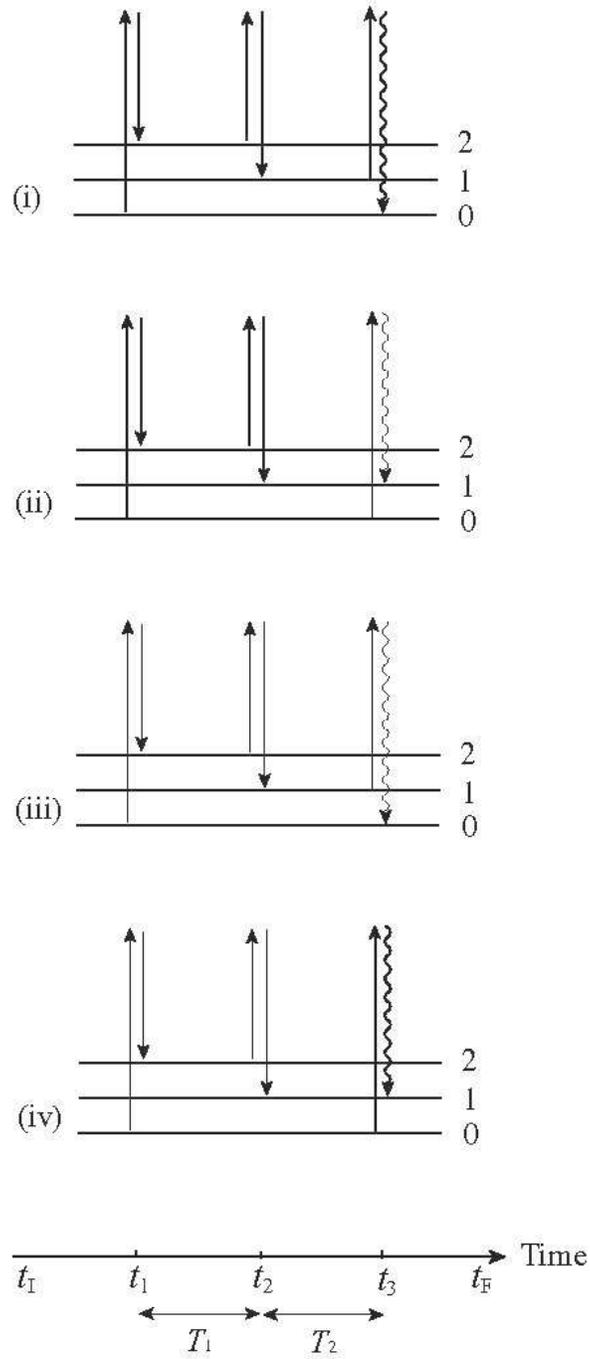}
\end{center}
\caption{Energy-level diagrams of $R^{(2)}(T_{1},T_{2})$ for Raman processes.
}%
\label{f1}%
\end{figure}

For the moment, we concentrate on the Raman case, i.e. $\left\langle \left[
\left[  \alpha\left(  t_{3}\right)  ,\alpha\left(  t_{2}\right)  \right]
,\alpha\left(  t_{1}\right)  \right]  \right\rangle $. Some of processes in
Eq. (\ref{R2}) are represented by the energy-level (Albrecht-like)
diagrams\ in Fig. \ref{f1}. The differences from the original Albrecht diagram
are mentioned at the end of this section. Before explaining diagrams, let us
review possible transitions by operators $Q$ and $Q^{2}$; $Q$ can cause a
one-quantum excitation or de-excitation while $Q^{2}$ can result in a
two-quantum excitation or de-excitation in addition to a zero-quantum
transition. For example, from $\left\vert 0\right\rangle \rightarrow
Q^{2}\left\vert 0\right\rangle \sim\left[  \left(  a^{\dag}\right)
^{2}+aa^{\dag}\right]  \left\vert 0\right\rangle $, we see that by the action
of the operator $Q^{2}$, the ground ket state $\left\vert 0\right\rangle $ can
be changed into $\left\vert 0\right\rangle $ (zero-quantum transition) or
$\left\vert 2\right\rangle $ (two-quantum excitation). In the same way,
$\left\langle 2\right\vert $ can be brought into $\left\langle 0\right\vert $
(two-quantum de-excitation) or $\left\langle 2\right\vert $.

In the diagrams, time runs from the left to the right. Each pair of arrows
stands for a Raman excitation. The pair with a wavy arrow signifies the Raman
induction decay (\emph{last} interaction); the first interaction occurs at
$t_{1}$, the second at $t_{2}$, and the last at $t_{3}$.

The full description of a quantum state at a certain time requires both the
bra state $\left\langle n\right\vert $ and ket state $\left\vert
m\right\rangle $; at any time the state is fully specified by the Liouville
state $\left\vert m\right\rangle \left\langle n\right\vert $. In the diagrams,
the excitation or de-excitation of the \emph{bra} state is expressed by a pair
of \emph{fine} arrows while that of the \emph{ket} state by \emph{normal}
ones. For example, the first interaction at $t_{1}$ of (i) and (ii) is a
two-quantum excitation of the ket state while that of (iii) and (iv) is of the
bra state.

In the Liouville space, the diagram (i) is interpreted as follows. The system
is initially in the ground (Liouville) state $\left\vert 0\right\rangle
\left\langle 0\right\vert $. The first interaction causes a two-quantum
excitation of the ket state; $\left\vert 0\right\rangle \left\langle
0\right\vert \rightarrow\left\vert 2\right\rangle \left\langle 0\right\vert $
at $t_{1}$. The second interaction causes a one-quantum de-excitation,
$\left\vert 2\right\rangle \left\langle 0\right\vert \rightarrow\left\vert
1\right\rangle \left\langle 0\right\vert $ at $t_{2}$. The last shows a
one-quantum de-excitation, $\left\vert 1\right\rangle \left\langle
0\right\vert \rightarrow\left\vert 0\right\rangle \left\langle 0\right\vert $
at $t_{3}$. As a whole, we denote this as%
\begin{equation}
\left\vert 0\right\rangle \left\langle 0\right\vert \underset{t_{1}%
}{\rightarrow}\left\vert 2\right\rangle \left\langle 0\right\vert
\underset{t_{2}}{\rightarrow}\left\vert 1\right\rangle \left\langle
0\right\vert \underset{t_{3}}{\rightarrow}\left\vert 0\right\rangle
\left\langle 0\right\vert \label{m1}%
\end{equation}

The diagram (ii)-(iv) are interpreted as follows:%
\begin{align}
&  \left\vert 0\right\rangle \left\langle 0\right\vert \underset{t_{1}%
}{\rightarrow}\left\vert 2\right\rangle \left\langle 0\right\vert
\underset{t_{2}}{\rightarrow}\left\vert 1\right\rangle \left\langle
0\right\vert \underset{t_{3}}{\rightarrow}\left\vert 1\right\rangle
\left\langle 1\right\vert \label{m2}\\
&  \left\vert 0\right\rangle \left\langle 0\right\vert \underset{t_{1}%
}{\rightarrow}\left\vert 0\right\rangle \left\langle 2\right\vert
\underset{t_{2}}{\rightarrow}\left\vert 0\right\rangle \left\langle
1\right\vert \underset{t_{3}}{\rightarrow}\left\vert 0\right\rangle
\left\langle 0\right\vert \label{m3}\\
&  \left\vert 0\right\rangle \left\langle 0\right\vert \underset{t_{1}%
}{\rightarrow}\left\vert 0\right\rangle \left\langle 2\right\vert
\underset{t_{2}}{\rightarrow}\left\vert 0\right\rangle \left\langle
1\right\vert \underset{t_{3}}{\rightarrow}\left\vert 1\right\rangle
\left\langle 1\right\vert \label{m4}%
\end{align}
Note here that a pair of \emph{fine} arrows always correspond to the
excitation or de-excitation of the \emph{bra} state.

We define the population state by $\left\vert n\right\rangle \left\langle
n\right\vert $, while the coherence state by $\left\vert n\right\rangle
\left\langle m\right\vert $ with $n\neq m$. We notice that, after the last
interaction, in all of the above 4 diagrams, the system is always in a
population state ($\left\vert 0\right\rangle \left\langle 0\right\vert $ or
$\left\vert 1\right\rangle \left\langle 1\right\vert $). In summary, \emph{a
diagram does not vanish only when the final state is a population state}
(Theorem 1). This corresponds to the trace operation in the definition of the
response function.

In this paper, we simplify the original Albrecht diagrams \cite{LeeAlbrecht}
for comparison with the Liouville paths. The main differences are the
following: (1) we use always the same horizontal lines regardless of ket or
bra states; it is not the case in the original Albrecht diagrams and (2) time
runs always from left to right in our representation while the direction for
the bra and ket states are the opposite in the original version. Our
representation is somewhat simpler in that a single diagram in ours
corresponding to several diagrams in the original version.

\subsection{IR and IR-Raman spectroscopy}

\begin{figure}
\begin{center}
\includegraphics[scale=0.7]{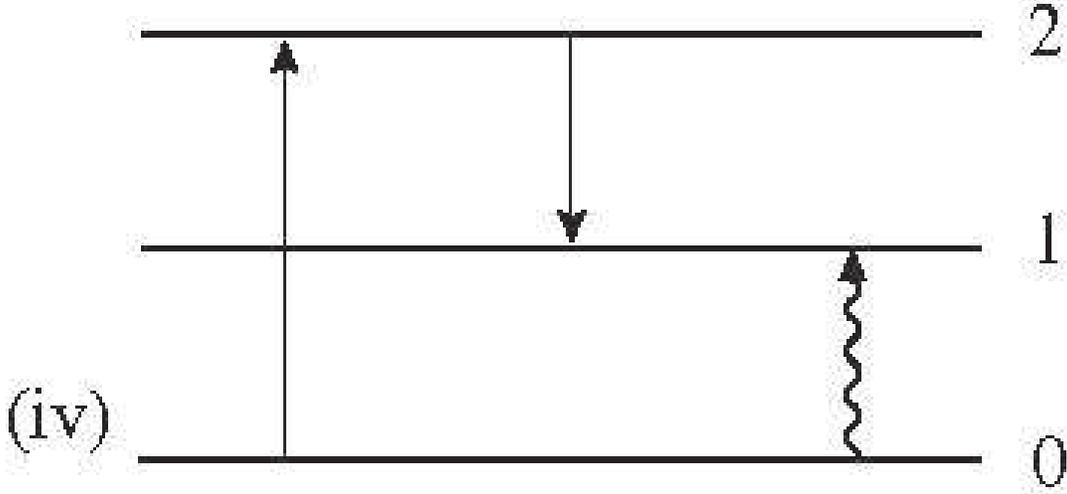}
\end{center}
\caption{An energy-level diagram of $R^{(2)}(T_{1},T_{2})$ for IR processes.}%
\label{f2}%
\end{figure}

IR processes appearing in the IR response function, $\left\langle \left[
\left[  \mu\left(  t_{3}\right)  ,\mu\left(  t_{2}\right)  \right]
,\mu\left(  t_{1}\right)  \right]  \right\rangle $, corresponding to Fig.
\ref{f1}-(iv) is described in Fig. \ref{f2}; each quantum transition is
represented not by a pair of arrow but an arrow. Note Raman and IR processes
can be equivalent theoretically at this level of description, although even
orders of IR processes, such as second-order IR signal vanish except in
anisotopic media, such as adsorbed molecules on metallic surface.
\cite{Chosurface} This situation can be overcome by mixing the IR and Raman
processes. \cite{PC} By using narrow-band lasers (two IR excitation pulses
followed one probe pulse which create Raman signal) Zhao and Wright
demonstrated such experiment. \cite{Wright,WrightA}

\begin{figure}
\begin{center}
\includegraphics[scale=0.7]{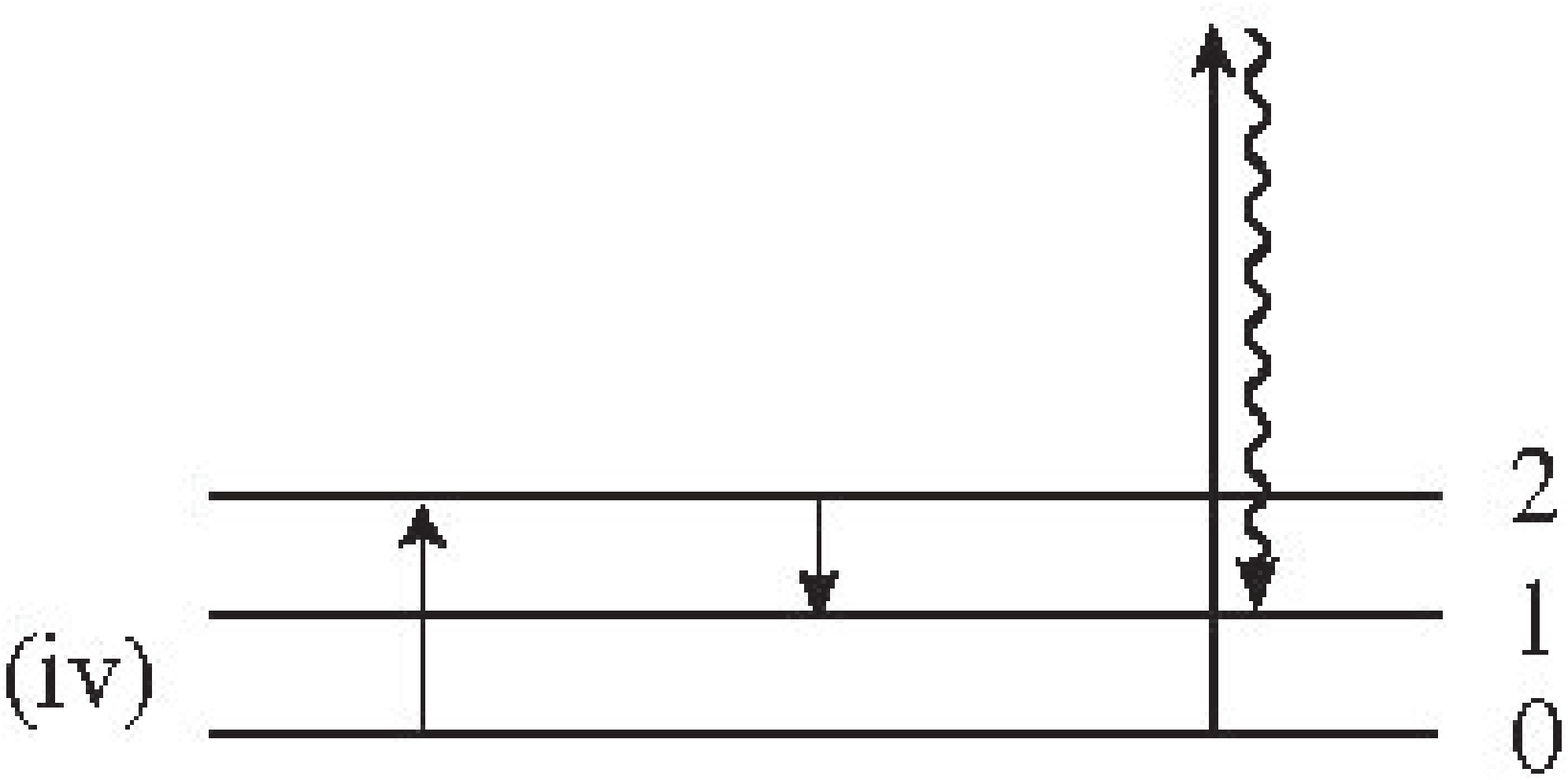}
\end{center}
\caption{An energy-level diagrams of $R^{(2)}(T_{1},T_{2})$ for IR-Raman
processes, $\left\langle \left[  \left[  \mu\left(  t_{3}\right)  ,\mu\left(
t_{2}\right)  \right]  ,\alpha\left(  t_{1}\right)  \right]  \right\rangle $.}%
\label{f3}%
\end{figure}

As an IR-Raman spectroscopy, we consider the response function, $\left\langle
\left[  \left[  \mu\left(  t_{3}\right)  ,\mu\left(  t_{2}\right)  \right]
,\alpha\left(  t_{1}\right)  \right]  \right\rangle $, for example. A diagram
corresponding to Fig.\ref{f1}-(iv) is shown in Fig. \ref{f3}; Raman and IR
transitions are represented by a pair of arrows and an arrow, respectively.
Diagrams corresponding to the other IR-Raman response function such as
$\left\langle \left[  \left[  \mu\left(  t_{3}\right)  ,\mu\left(
t_{2}\right)  \right]  ,\alpha\left(  t_{1}\right)  \right]  \right\rangle $
can be described in a similar manner.

\section{Energy-level diagram and double-sided diagram}

\begin{figure}
\begin{center}
\includegraphics[scale=0.6]{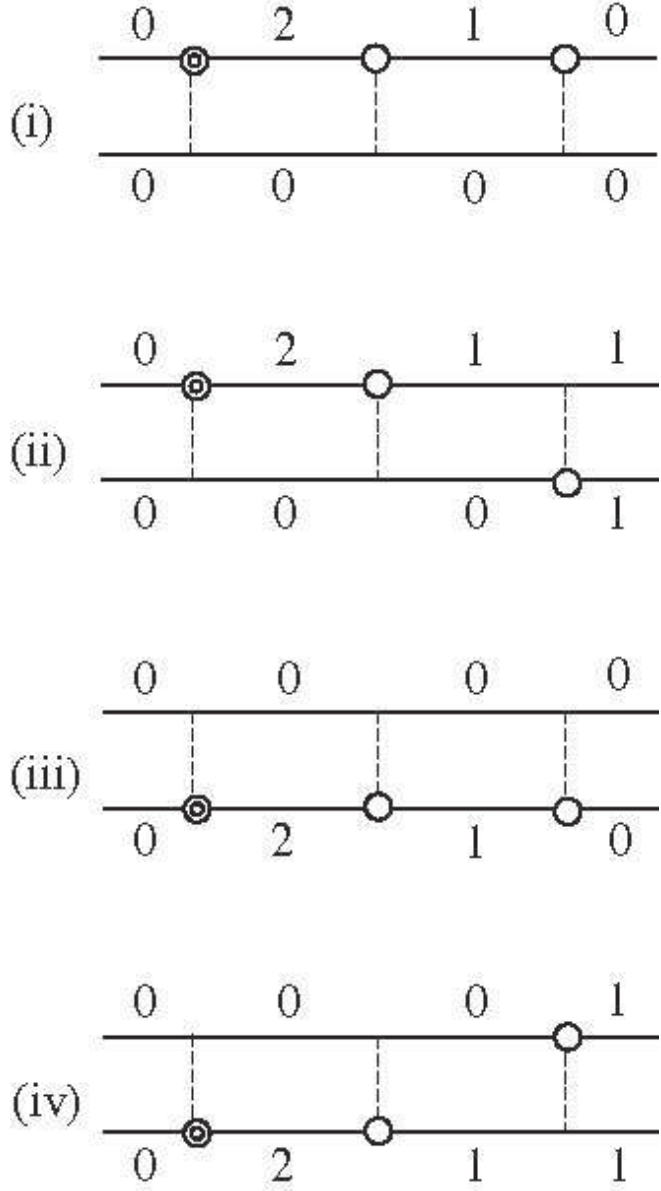}
\end{center}
\caption{Double-sided Feynman diagrams of $R^{(2)}(T_{1},T_{2})$.}%
\label{f4}%
\end{figure}

We can represent processes in the Liouville-space in a different way by the
double-sided Feynman diagrams. The diagrams in Fig. \ref{f4} are the
translation of the diagrams in Fig. \ref{f1}, \ref{f2}, or \ref{f3}. In the
double-sided diagrams, time runs from the left to the right (as in the
energy-level diagram). The horizontal lines, however, are always two in
number, the upper and lower line. The former represents the ket state while
the latter the bra state. The single circle stands for a one-quantum
transition, while the double circle for a two quantum transition. The quantum
number of the bra and ket states is denoted explicitly in the diagram.

It is noted that there are some differences of diagrammatic notation among
articles. For example, in some literature, the quantum transition is not
represented by \ circles but arrows. In other one, diagrams are rotated by 90
degree so that the time runs from the bottom to the top.

In general, as seen below (VI. A), \emph{the double-sided diagram is
convenient for enumeration of all possible diagrams while the energy-level
diagram is for understanding the physical process}.

\section{Feynman rules for the diagrams}

We have introduced several way to represent optical processes as in Figs.
\ref{f1}-\ref{f4}. It is emphasized here that the interpretation in terms of
the Liouville-space state $\left\vert m\right\rangle \left\langle n\right\vert
$ is unique except for what $x_{k}$ implies. Accordingly, we can develop a
universal rule to write down analytical expressions from diagrams via the
interpretations (such as Eqs. (\ref{m1})-(\ref{m4})) in the Liouville-space;
the derivation is a straightforward exercise in elementary quantum mechanics
and would be discussed elsewhere. It can be summarized in the following way.
We associate with each interaction (originating from the interaction
$Q^{k}/k!$) at a certain time or each propagation for a certain period one of
the following factors:

\begin{center}%
\begin{tabular}
[c]{c||c}%
$\text{interaction (}n\geq0\text{)}$ & $\text{factor}$\\\hline\hline
$\left\vert m\right\rangle \rightarrow\left\vert m+n\right\rangle $ &
$\frac{i}{\hbar}x_{k}\left\langle m+n\right\vert Q^{k}\left\vert
m\right\rangle /k!$\\
$\left\vert m\right\rangle \rightarrow\left\vert m-n\right\rangle $ &
$\frac{i}{\hbar}x_{k}\left\langle m-n\right\vert Q^{k}\left\vert
m\right\rangle /k!$\\
$\left\langle m\right\vert \rightarrow\left\langle m+n\right\vert $ &
$-\frac{i}{\hbar}x_{k}\left\langle m\right\vert Q^{k}\left\vert
m+n\right\rangle /k!$\\
$\left\langle m\right\vert \rightarrow\left\langle m-n\right\vert $ &
$-\frac{i}{\hbar}x_{k}\left\langle m\right\vert Q^{k}\left\vert
m-n\right\rangle /k!$\\
$\text{remark}$ & $\text{omit }\pm\frac{i}{\hbar}\text{ for the last
interaction}$%
\end{tabular}

\smallskip\bigskip%

\begin{tabular}
[c]{c||c}%
pr$\text{opagation (}t\geq0\text{)}$ & f$\text{actor}$\\\hline\hline
$\text{ }\left\vert m\right\rangle \left\langle n\right\vert \text{ for }t$ &
$e^{-i\zeta_{mn}t-\Gamma_{mn}t}\text{ }$%
\end{tabular}

\end{center}

By multiplying all the factors and putting another factor 1/2 to avoid
double-counting (see Theorem 2 below), we obtain an analytical expression of
the corresponding diagram (\emph{Feynman rule}). Here, we have introduced
$\zeta_{mn}$ and $\Gamma_{mn}$ $(\geq0)$ to describe relaxation; the
difference of frequency modified due to the relaxation is defined by
$\zeta_{mn}=(m-n)\zeta$ while the relaxation constant $\Gamma_{mn}$ for the
state $\left\vert m\right\rangle \left\langle n\right\vert $ possesses the
symmetric property, $\Gamma_{nm}=\Gamma_{mn},$ which is a necessary condition
for a consistent theory (see below Eq. (\ref{Vi})). Without dissipation,
$\zeta_{mn}\rightarrow\Omega_{mn}=(m-n)\Omega$ and $\Gamma_{mn}\rightarrow0.$
In the Brownian oscillator model with the damping constant $\gamma$, the
corrected frequency $\zeta$ is given by $\zeta=\sqrt{\Omega^{2}-\left(
\gamma/2\right)  ^{2}}$. \cite{Grabert,Weiss} The expression for $\Gamma_{mn}$
in this model shall be discussed below.

By definition, \emph{the propagation period} implies the time \emph{between
two interactions}. This excludes the periods from $t_{I}$ to $t_{1}$ and from
$t_{3}$ to $t_{F}$ in diagrams in Figs. \ref{f1}-\ref{f4} (or, say, in Eq.
(\ref{m1})-(\ref{m4})) because there is no interaction at $t_{I}$ or $t_{F}$;
we associate the unity for these special period.

Let us apply our rule without relaxation ($\Gamma_{mn}=0$, $\zeta_{mn}%
=\Omega_{mn}$) to a diagram or a Liouville-space path. As the first example,
we consider the diagram (i) (of Fig. \ref{f1} or \ref{f4}). We have only 2
separate propagation periods by definition. In the first period from $t_{1}$
to $t_{2}$ the system is in the state $\left\vert 2\right\rangle \left\langle
0\right\vert $ and thus we have the factor $e^{-i\Omega_{20}\left(
t_{2}-t_{1}\right)  }$ while for the last period from $t_{2}$ to $t_{3}$ the
system is in the state $\left\vert 1\right\rangle \left\langle 0\right\vert $
and we have the factor $e^{-i\Omega_{10}\left(  t_{3}-t_{2}\right)  }$; in
total we have the propagation factor, $e^{-i\Omega_{20}T_{1}}\cdot
e^{-i\Omega_{10}T_{2}}$, where we have used the relation (\ref{td}). In
addition, as the result of the three interactions, we have other factors,
$\frac{i}{\hbar}x_{2}\left\langle 2\right\vert Q^{2}\left\vert 0\right\rangle
/2\cdot\frac{i}{\hbar}x_{1}\left\langle 1\right\vert Q\left\vert
2\right\rangle \cdot x_{1}\left\langle 0\right\vert Q\left\vert 1\right\rangle
=\left(  \frac{i}{\hbar}\frac{\hbar}{2M\Omega}x_{1}\right)  ^{2}x_{2}$ (Note
here the relations, Eq. (\ref{Q}) as well as, $a\left\vert n\right\rangle
=\sqrt{n}\left\vert n-1\right\rangle ,$ and $a^{\dag}\left\vert n\right\rangle
=\sqrt{n+1}\left\vert n+1\right\rangle $). In summary, the process in Eq.
(\ref{m1}) or the diagram (i) is given (with the extra factor 1/2 associate
with double-counting) by
\begin{equation}
(\text{i})=\left(  \frac{i}{\hbar}\right)  ^{2}\frac{x_{1}^{2}x_{2}}{2}\left(
\frac{\hbar}{2M\Omega}\right)  ^{2}e^{-i2\Omega T_{1}}\cdot e^{-i\Omega T_{2}}
\label{i}%
\end{equation}

The process in Eq. (\ref{m2}) or the diagram (ii) (of Figs. \ref{f1} or
\ref{f4}) is different from (i) only after $t_{3}$. Although the last
interaction at $t_{3}$ is that for the bra state (expressed by the fine arrows
and different form (i)) the factors for this last interaction is the same with
that of (i) by the above Feynman rule; there is no sign differences between
bra and ket states (only) for the last interaction. In summary we have%
\[
(\text{ii})=(\text{i}).
\]
In general, we have the following theorem, which is related to the double
counting: \emph{The diagrams different only by the side of the last
interaction (bra or ket side) have the same contribution} (Theorem2).

The process in Eq. (\ref{m3}) or in the diagram (iii) can be estimated in a
similar manner by the above Feynman rule:%
\[
(\text{iii})=\left(  -\frac{i}{\hbar}\right)  ^{2}\frac{x_{1}^{2}x_{2}}%
{2}\left(  \frac{\hbar}{2M\Omega}\right)  ^{2}e^{i2\Omega T_{1}}\cdot
e^{i\Omega T_{2}}.
\]
Note here that the sign in front of $i/\hbar$ is minus because of the
interactions on the bra state (fine arrows). From $t_{1}$ to $t_{2}$, the
system is in the states $\left\vert 0\right\rangle \left\langle 2\right\vert $
and $\left\vert 2\right\rangle \left\langle 0\right\vert $ in (iii) and (i),
respectively; these two states are the complex-conjugate of each other. From
$t_{2}$ to $t_{3}$, the state of (iii) ($\left\vert 0\right\rangle
\left\langle 1\right\vert $) is again in the complex-conjugate state of (i)
($\left\vert 1\right\rangle \left\langle 0\right\vert $). Accordingly, (iii)
given in the above is the complex conjugate of (i), i.e.,$($iii$)=($i$)^{\ast
}$. Diagrammatically, in (iii) of Fig. \ref{f1}, all the normal arrows in (i)
are replaced by the fine arrows. In general, \emph{The complex-conjugate
diagram is obtained by interchanging all the normal and fine arrows }(Theorem
3). In the double-sided Feynman diagrams, instead, \emph{The complex-conjugate
diagram is obtained by interchanging the circles on the upper and lower lines
}(Theorem 3$^{\prime}$).

The diagram (iv) is the complex-conjugate diagram of (ii) because the fine and
normal arrows are interchanged, i.e., $($iv$)=($ii$)^{\ast}.$ We can also
verify the relation, $($iii$)=($iv$)$, from the above Feynman rule with
reconfirming Theorem 2.

\section{Temperature effect and initial state}

In the above, we have assumed the system is initially in the ground state
$\left\vert 0\right\rangle \left\langle 0\right\vert $, which is usually
justified for high frequency vibration modes at a room temperature. For high
temperatures or low frequency modes, however, excited states $\left\vert
n\right\rangle \left\langle n\right\vert $ are initially populated according
to the Boltzmann factor. In general, we have to estimate all the possible
processes assuming that the system is initially in the population state
$\left\vert n\right\rangle \left\langle n\right\vert $ using the above
mentioned rule, and then summing up with respect to $n$ with the Boltzmann
factor $e^{-\beta E_{n}}/\sum_{n}e^{-\beta E_{n}}$ (in the case without
dissipation); this completes our Feynman rule.

\begin{figure}
\begin{center}
\includegraphics[scale=0.6]{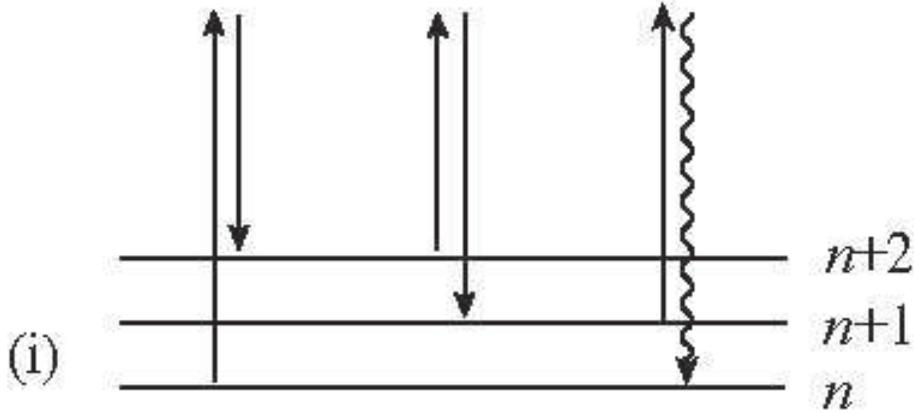}
\end{center}
\caption{General process corresponding to Fig. \ref{f1} (i)}%
\label{f6}%
\end{figure}

Even if we take into account the contribution from general initial state
$\left\vert n\right\rangle \left\langle n\right\vert $, however, in the
(fully-corrected) Ohmic Brownian oscillator model, we still have the same
result with above as shown in the previous literature. This is the reflection
of the relation%
\[
\left\langle n\right\vert X\left\vert n\right\rangle =\left\langle
0\right\vert X\left\vert 0\right\rangle
\]
where $X$ is some special combination of operators (This could be directly
checked by laborious calculation by using our Feynman rule). The fact that
$R^{(2)}(T_{1},T_{2})$ treated in this paper is independent of the temperature
and thus we can obtain a finite temperature result even if assuming that the
system is initially in the ground state is by no means trivial but established
by other calculation methods. \cite{OT} This implies, for example, that the
dependence on $n$ of the analytical expression corresponding to Fig. \ref{f6}
cancels out with some other diagram. When the damping mechanism other than
(fully-corrected) Ohmic Brownian oscillator model, our results presented below
might be interpreted as an high frequency approximation, i.e. $\hbar\Omega\gg
kT$. 

\section{Liouville-space quartet}

The four diagrams (i)-(iv) in Figs. \ref{f1} and \ref{f4} are a special set in
the sense that we can obtain the other three, starting from one of the quartet.

In the energy-level diagram, we obtain the second by changing the last
interaction by using one of the following rule (depending on the last
interaction of the starting diagram); (1) the ket excitation to a bra
de-excitation, (2) the ket de-excitation to a bra excitation, (3) the bra
excitation to a ket de-excitation and (4) the bra de-excitation to a ket
excitation. The remaining two diagrams are the complex-conjugate diagrams of
the previous two diagrams where the conjugates are obtained by interchanging
the fine and normal arrows.

In the double-sided diagram the second diagram is obtained by lowering or
raising the last circle. The remaining two is by interchanging lower and upper
line with circles.

As seen before, the corresponding analytical expressions of (i)-(iv) have the
relations, $($i$)=($ii$),$ $($iii$)=($iv$),$ and $($i$)=($iii$)^{\ast}.$ The
sum of the quartet is always real:%
\[
(\text{i})+(\text{ii})+(\text{iii})+(\text{iv})=4\operatorname{Re}%
[(\text{i})]=4\operatorname{Re}[(n)]
\]
where $n=$ i, ii, iii or iv. Taking the real part of Eq. (\ref{i}) \ we have
an expression for the quartet,
\begin{equation}
V_{I}=-\frac{x_{1}^{2}x_{2}}{2\left(  M\Omega\right)  ^{2}}\cos\left(  2\Omega
T_{1}+\Omega T_{2}\right)  \label{Vi}%
\end{equation}

In terms of the interpretation in the Liouville space in Eqs. (\ref{m1}%
)-(\ref{m4}), all the processes posses a common property; the two-quantum
coherence ($\left\vert 2\right\rangle \left\langle 0\right\vert $ or
$\left\vert 0\right\rangle \left\langle 2\right\vert $) is realized for
$T_{1}$ while the one-quantum coherence ($\left\vert 1\right\rangle
\left\langle 0\right\vert $ or $\left\vert 0\right\rangle \left\langle
1\right\vert $) for $T_{2}$; we denote this as:%
\[
\left\vert 2\right\rangle \left\langle 0\right\vert \rightarrow\left\vert
1\right\rangle \left\langle 0\right\vert \text{ and }\left\vert 0\right\rangle
\left\langle 2\right\vert \rightarrow\left\vert 0\right\rangle \left\langle
1\right\vert
\]
This is reflected by the factor $\cos\left(  2\Omega T_{1}+\Omega
T_{2}\right)  $ in Eq. (\ref{Vi}).

We notice that in the case with damping if $\Gamma_{mn}$ (and $\zeta_{mn}$)
were not symmetric, $V_{I}$ could not be real; the symmetric property of
$\Gamma_{mn}$ is required for the response function to be real.

\subsection{Quartets representation: all possible quartets for $R^{(2)}%
(T_{1},T_{2})$}

\begin{figure}
\begin{center}
\includegraphics[scale=0.45]{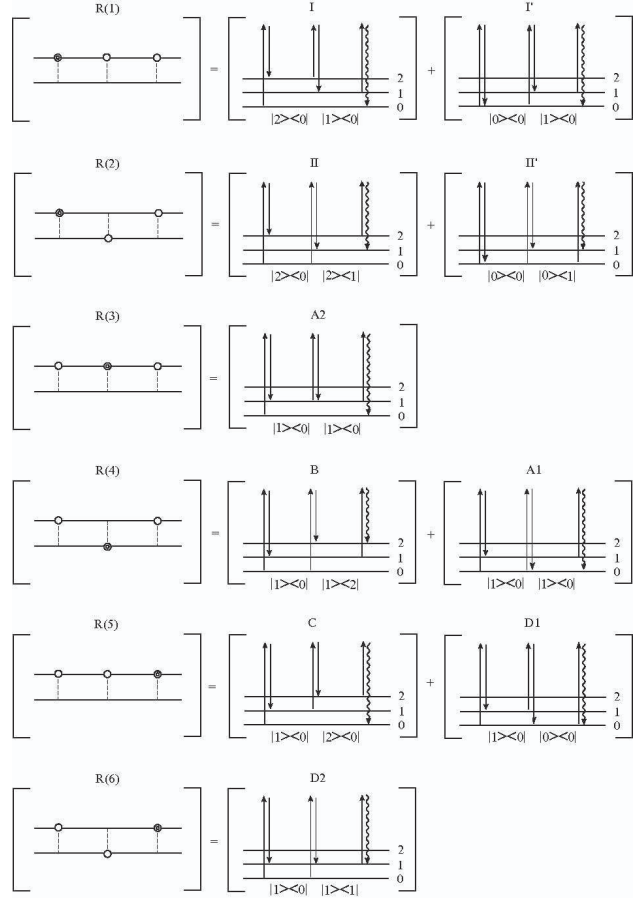}
\end{center}
\caption{All possible quartets for $R^{(2)}(T_{1},T_{2})$. The square bracket
implies that four diagrams are collectively represented. For example, the
first diagram in the energy-level diagram for R(1) corresponds to not only (i)
of Fig. \ref{f1} (which is explicitly written in the bracket) but other three
diagrams (ii)-(iv) of Fig. \ref{f1}.}%
\label{f7}%
\end{figure}

We show six quartets R(1)$-$R(6) in Fig. \ref{f7} in the double-sided
representation. The square brackets imply the quartet; only one of the quartet
is explicitly written in the bracket. For example, R(1) of Fig. \ref{f7}
collectively stands for (i)-(iv) of Fig. \ref{f4}.

In Fig. \ref{f7}, on the right side, ten quartets in the energy-level
representation are given; some quartets in double-sided representation
corresponds to not one but two quartets in the energy-level representation.
For example, R(1) contains contribution I and I$^{\prime}$, while R(3)
contains only A2.

These six quartets R(1)$-$R(6) in Fig. \ref{f7} exhaust all
possible contribution to the right-hand side of Eq. (\ref{R2}); there are 3
ways for the\ position of the double quantum transition (double circle) and
there $2^{3}$ ways to put the three (including one double circle) circles
upper or lower line, which leads to $3\cdot2^{3}$ double-sided Feynman
diagrams in total. These $3\cdot8$ diagrams can be divided into $6$ quartets
that have been shown. We understand here that \emph{the double-sided diagram
is convenient for enumerating all possible diagrams}.

\subsection{Estimation of quartets}

The analytical expression of quartet II is given via our Feynman rule:
\begin{align}
\text{II}  &  =4\operatorname{Re}\left[  -\left(  \frac{i}{\hbar}\right)
^{2}\left(  \frac{\hbar}{2M\zeta}\right)  ^{2}\right. \label{ii}\\
&  \times\left.  \frac{x_{1}^{2}x_{2}}{2}\cdot e^{-i2\zeta T_{1}-\Gamma
_{20}T_{1}}\cdot e^{-i\zeta T_{2}-\Gamma_{21}T_{2}}\right]  ,
\end{align}
where the analytical expression in the square bracket has been derived from
the diagram explicitly drawn in the bracket in Fig. \ref{f7} (in the presence
of dissipation). For example, the propagator $e^{-i2\zeta T_{1}-\Gamma
_{20}T_{1}}$ and $e^{-i\zeta T_{2}-\Gamma_{21}T_{2}}$ come from the
propagation of $\left\vert 2\right\rangle \left\langle 0\right\vert $ and
$\left\vert 2\right\rangle \left\langle 1\right\vert $, respectively.

In this way we obtain the expression:%
\begin{equation}
R^{(2)}\left(  T_{1},T_{2}\right)  =\text{I}+\text{II}+\text{A}+\text{B}%
+\text{C}+\text{D1+D2} \label{r2}%
\end{equation}
with%
\begin{align*}
\text{I}  &  =-\frac{x_{1}^{2}x_{2}}{2\left(  M\zeta\right)  ^{2}}%
e^{-\Gamma_{20}T_{1}-\Gamma_{10}T_{2}}\cos\left(  2\zeta T_{1}+\zeta
T_{2}\right) \\
\text{II}  &  =\frac{x_{1}^{2}x_{2}}{2\left(  M\zeta\right)  ^{2}}%
e^{-\Gamma_{20}T_{1}-\Gamma_{21}T_{2}}\cos\left(  2\zeta T_{1}+\zeta
T_{2}\right) \\
\text{A}  &  =-\frac{x_{1}^{2}x_{2}}{2\left(  M\zeta\right)  ^{2}}%
e^{-\Gamma_{10}T_{1}-\Gamma_{10}T_{2}}\cos\left(  \zeta T_{1}+\zeta
T_{2}\right) \\
\text{B}  &  =\frac{x_{1}^{2}x_{2}}{2\left(  M\zeta\right)  ^{2}}%
e^{-\Gamma_{10}T_{1}-\Gamma_{12}T_{2}}\cos\left(  \zeta T_{1}-\zeta
T_{2}\right) \\
\text{C}  &  =-\frac{x_{1}^{2}x_{2}}{2\left(  M\zeta\right)  ^{2}}%
e^{-\Gamma_{10}T_{1}-\Gamma_{20}T_{2}}\cos\left(  \zeta T_{1}+2\zeta
T_{2}\right) \\
\text{D1}  &  =-\frac{1}{4}\frac{x_{1}^{2}x_{2}}{\left(  M\zeta\right)  ^{2}%
}e^{-\Gamma_{10}T_{1}-\Gamma_{00}T_{2}}\cos\left(  \zeta T_{1}\right) \\
\text{D2}  &  =\frac{3}{4}\frac{x_{1}^{2}x_{2}}{\left(  M\zeta\right)  ^{2}%
}e^{-\Gamma_{10}T_{1}-\Gamma_{11}T_{2}}\cos\left(  \zeta T_{1}\right)
\end{align*}
As for the derivation of this we remark: (1) Quartets I' and II' cancel out
because I'$=-\frac{1}{4}\frac{x_{1}^{2}x_{2}}{\left(  M\zeta\right)  ^{2}%
}e^{-\Gamma_{00}T_{1}-\Gamma_{10}T_{2}}\cos\left(  \zeta T_{2}\right)  $ and
II'$=\frac{1}{4}\frac{x_{1}^{2}x_{2}}{\left(  M\zeta\right)  ^{2}}%
e^{-\Gamma_{00}T_{1}-\Gamma_{01}T_{2}}\cos\left(  \zeta T_{2}\right)  $ (The
numerical factor 1/4 can be understood from the first two-quantum transition
associated with $\left\langle 0\right\vert Q^{2}\left\vert 0\right\rangle
\varpropto\left\langle 0\right\vert aa^{\dag}\left\vert 0\right\rangle =1$).
(2) The sum A2+A1 reduces to A (The numerical factor for A2 (or A1) can be
estimated by noting the second two-quantum transition $\left\langle
1\right\vert Q^{2}\left\vert 1\right\rangle \varpropto\left\langle
1\right\vert aa^{\dag}+a^{\dag}a\left\vert 1\right\rangle =3$ (or
$\left\langle 0\right\vert Q^{2}\left\vert 0\right\rangle \varpropto
\left\langle 0\right\vert aa^{\dag}\left\vert 0\right\rangle =1$)).

It is worth while observing the relationships between analytical expressions
and the symbolic interpretations of the remaining quartets:%
\begin{align*}
\text{A}  &  \text{: }\left\vert 1\right\rangle \left\langle 0\right\vert
\rightarrow\left\vert 1\right\rangle \left\langle 0\right\vert \text{ and
}\left\vert 0\right\rangle \left\langle 1\right\vert \rightarrow\left\vert
0\right\rangle \left\langle 1\right\vert \\
\text{B}  &  \text{: }\left\vert 1\right\rangle \left\langle 0\right\vert
\rightarrow\left\vert 1\right\rangle \left\langle 2\right\vert \text{ and
}\left\vert 0\right\rangle \left\langle 1\right\vert \rightarrow\left\vert
2\right\rangle \left\langle 1\right\vert \\
\text{C}  &  \text{: }\left\vert 1\right\rangle \left\langle 0\right\vert
\rightarrow\left\vert 2\right\rangle \left\langle 0\right\vert \text{ and
}\left\vert 0\right\rangle \left\langle 1\right\vert \rightarrow\left\vert
0\right\rangle \left\langle 2\right\vert \\
\text{D2}  &  \text{: }\left\vert 1\right\rangle \left\langle 0\right\vert
\rightarrow\left\vert 1\right\rangle \left\langle 1\right\vert \text{ and
}\left\vert 0\right\rangle \left\langle 1\right\vert \rightarrow\left\vert
1\right\rangle \left\langle 1\right\vert \\
\text{D1}  &  \text{: }\left\vert 1\right\rangle \left\langle 0\right\vert
\rightarrow\left\vert 0\right\rangle \left\langle 0\right\vert \text{ and
}\left\vert 0\right\rangle \left\langle 1\right\vert \rightarrow\left\vert
0\right\rangle \left\langle 0\right\vert
\end{align*}
That is, we can associate the state $\left\vert n\right\rangle \left\langle
m\right\vert $ with $\zeta_{n,m}$\thinspace and $\Gamma_{mn}$.

In addition, if we fully include the temperature effect by our Feynman rule
with tracking all the possible processes, we could obtain the result given in
Appendix B of \cite{Steffen}.

\section{Damping mechanism}

We can confirm that the well-known result for the Ohmic Brownian oscillator
(BO) model (Ohmic implies that the system-bath coupling is in the bilinear
form) is reproduced from Eq. (\ref{r2}) by setting%
\begin{equation}
\text{ }\Gamma_{nm}=\left\{
\begin{array}
[c]{cc}%
\gamma & \text{for }\left\vert n\right\rangle \left\langle n\right\vert \\
\left\vert n-m\right\vert \gamma/2 & \text{for }\left\vert n\right\rangle
\left\langle m\right\vert \text{ (}n\neq m\text{)}%
\end{array}
\right.  \text{ } \label{relaxB}%
\end{equation}
where $\left\vert m\right\vert $ represents the absolute value of $m$.
Actually, in the Brownian result, I+II should be zero, which is true if
$\Gamma_{21}=\Gamma_{10}$, while D1+D2 should be $-2\cdot$D1, which is true if
$\Gamma_{11}=\Gamma_{00}$; $\Gamma_{mn}$ in Eq. (\ref{relaxB}) satisfies these requirements.

The cancellation of I and II is one of the feature of the Brownian result.
Another feature is that the state $\left\vert 0\right\rangle \left\langle
0\right\vert $ decays with the relaxation constant $\gamma/2$ which is the
same as that for $\left\vert 1\right\rangle \left\langle 1\right\vert $. These
characteristics have intrigued some controversy as mentioned below.

The relaxation constant for the same Ohmic model within the lower level
approximation, i.e., at the level of the Fermi's golden rule with a somewhat
ad hoc approximation (see below), given by \cite{Steffen,Fourkas95}
\begin{equation}
\Gamma_{mn}=\frac{n+m}{2}\gamma, \label{relaxF}%
\end{equation}
which is also simple but incompatible with the above two requirements
($\Gamma_{21}=\Gamma_{10}$ and $\Gamma_{11}=\Gamma_{00}$). With this
relaxation constant, I and II survives, for example. (In addition, there is no
frequency shift ($\zeta_{mn}\rightarrow\Omega_{mn}$) in this finite-order approximation).

The frequency shift and appearance of the absolute value ($\left\vert
n-m\right\vert $), which is \emph{non-analytic}, in the off diagonal
relaxation constant in the fully corrected expressions originate from the
summation of infinite number of diagrams; in the well-known result of Ohmic BO
model the bilinear coupling between \emph{the system-bath is fully taken into
account}; this is the exact prediction from a simple reasonable model and we
concern with the relaxation of \emph{fully-dressed states} in the exact result
of BO model. On the contrary, the relaxation constant in Eq. (\ref{relaxF}),
is the result of the same model but with the second-order (in the coupling
strength) approximation. Nonetheless in some context the second-order result
has been favored while the full-order result has been questioned.
\cite{Cho,Steffen}

As we show below we can distinguish the above two models ((\ref{relaxB}) or
(\ref{relaxF})) by some two-dimensional experiment by checking existence or
absence of certain peaks. In other words, whether the coherence (off-diagonal)
relaxation constant which depends only on the quantum number difference (where
$\Gamma_{m+n,m}=\Gamma_{n,0}$) and the level independent population relaxation
is appropriate (as the first-order picture) or not might be checked experimentally.

Note that if the system has some sort of anharmonicity such as the
anharmonicity of potential \cite{Oxtoby} or the nonlinear system-bath coupling
\cite{KT}, the relaxation constants do not hold the relation $\Gamma
_{21}=\Gamma_{10}$ etc., even we take into account higher-order system bath
interactions. Then the number of Liouville paths involved in the optical
processes increase dramatically especially when the system-bath interaction is
very strong. Also if the laser-molecular interaction is much shorter than the
time duration of the system-bath interactions, one has to regard the
relaxation rate as a function of time, i.e. $\Gamma_{nm}(t)$. In such case,
the equation of motion approach is more appropriate than the diagrammatic
approaches, although it requires computationally expensive calculations.
\cite{40,T,ST,TS} \cite{KT} \cite{KTnew}

We comment on confusion in the literature with regard to the Redfield theory,
one example of which is Eq. (\ref{relaxF}). The Redfield theory without the
rotational wave approximation (RWA) is equivalent to the Fokker-Planck
equation. \cite{40,T,ST,TS,KT,KTnew} The time evolution operator in the
Liouville space from the state $\left\vert \left.  k,l\right\rangle
\right\rangle \equiv\left\vert k\right\rangle \left\langle l\right\vert $ to
$\left\langle \left\langle i,j\right.  \right\vert \equiv$ $\left\langle
i\right\vert \cdots\left\vert j\right\rangle $ is then expressed as
$\left\langle \left\langle i,j\right.  \right\vert e^{-i\left(  \hat
{H}^{\times}-\hat{\Gamma}\right)  t}\left\vert \left.  k,l\right\rangle
\right\rangle $, where $\hat{H}^{\times}$ is the quantum Liouvillian and
$\hat{\Gamma}$ is the damping operator (Redfield operator). In energy-level
representation, $\left\vert \left.  k,l\right\rangle \right\rangle $ is the
eigen-function of the Hamiltonian but not the eigen function of $\hat{\Gamma}%
$, which makes difficult to evaluate this propagator. However, one sometimes
\textquotedblleft reads off\textquotedblright\ the damping constant directly
from the Redfield tensor elements $\Gamma_{ijkl}$ and incorporate them in the
propagator as $\left\langle \left\langle i,j\right.  \right\vert e^{-i\left(
\hat{H}^{\times}-\Gamma_{ijkl}\right)  t}\left\vert \left.  k,l\right\rangle
\right\rangle $, which can not be justified from the coordinate representation
model. \cite{Steffen,Fourkas95} Accordingly, this ad hoc methodology possesses
a flaw in the sense that the theory thus obtained does not converges to
analytical perturbative results such as obtained by the Brownian oscillator
model. It is possible to evaluate effective tensor element $\Gamma
_{ijkl}^{(\text{eff})}$ by solving the equation of motion such as the
Fokker-Planck equation with linear and nonlinear system-bath interactions,
\cite{40,T,ST,TS,KT,KTnew} but the calculated results are quite different from
the Redfield tensor elements. \cite{KT}

\section{Multi-mode system}

Extension to the multi-mode system, whose characteristic modes are represented
by $\left\{  Q_{s}\right\}  $, $\left\{  M_{s}\right\}  $, and $\left\{
\gamma_{s}\right\}  $, is straightforward. \cite{CPLfreq,TokPRL}%
\cite{Tominaga01,Tok01} We expand the dipole or polarizability operator as%
\[
x=x_{0}+\sum_{s}x_{1}^{(s)}Q^{(s)}+\frac{1}{2!}\sum_{s,s^{\prime}}%
x_{2}^{(ss^{\prime})}Q^{(s)}Q^{(s^{\prime})}+\cdots,
\]
and we denote the Liouville state by
\[
\left\vert \left\{  n_{s}\right\}  \right\rangle \left\langle \left\{
n^{\prime}\right\}  \right\vert =\left(  \left\vert n_{1}\right\rangle
\left\langle n_{1}^{\prime}\right\vert \right)  _{1}\cdots\left(  \left\vert
n_{s}\right\rangle \left\langle n_{s}^{\prime}\right\vert \right)  _{s}%
\cdots,
\]
where $\left\{  n_{s}\right\}  =(n_{1},n_{2},\cdots)$ is the quantum-number of
the corresponding mode. Here and hereafter, we use the notation in which
$\left\vert n_{s},n_{s^{\prime}}\right\rangle \left\langle m_{s},m_{s^{\prime
}}\right\vert $ stands for the state where the mode $s$ and $s^{\prime}$ are
in the states $\left\vert n_{s}\right\rangle \left\langle m_{s}\right\vert $
and $\left\vert n_{s^{\prime}}\right\rangle \left\langle m_{s^{\prime}%
}\right\vert $, respectively. For example, $\left\vert 0,1\right\rangle
\left\langle 2,3\right\vert $ means that the first and the second modes are in
the ground and the first excited ket states while they are in the second and
the third excited bra state, respectively.

The factor (in the Feynman rule) for the transition is well explained by
example. The transition,%
\[
\left\vert 0,0\right\rangle \left\langle 0,0\right\vert \rightarrow\left\vert
2,1\right\rangle \left\langle 0,0\right\vert
\]
caused by the operator $\left(  Q^{(1)}\right)  ^{2}Q^{(2)}$ is associated
with the factor $\frac{i}{\hbar}\left(  x_{3}^{(112)}+x_{3}^{(121)}%
+x_{3}^{(211)}\right)  \left\langle 2,1\right\vert \left(  Q^{(1)}\right)
^{2}Q^{(2)}\left\vert 0,0\right\rangle /3!=\frac{i}{\hbar}x_{3}^{(112)}%
\sqrt{2}\frac{\hbar}{M_{1}\Omega_{1}}\sqrt{\frac{\hbar}{M_{2}\Omega_{2}}}/2!$
while the transition (again caused by $\left(  Q^{(1)}\right)  ^{2}Q^{(2)}$),%
\[
\left\vert 0,0\right\rangle \left\langle 0,0\right\vert \rightarrow\left\vert
0,0\right\rangle \left\langle 2,1\right\vert
\]
is associated with the same factor with the minus sign. If the above
transition occurs at the \emph{last time}, however, we have to omit the factor
$i/\hbar$ as in the single-mode case.

Note here that the transition of the type,%
\[
\left|  0,0\right\rangle \left\langle 0,0\right|  \rightarrow\left|
1,0\right\rangle \left\langle 1,0\right|
\]
cannot occur at once, but
\[
\left|  0,0\right\rangle \left\langle 0,0\right|  \rightarrow\left|
1,1\right\rangle \left\langle 0,0\right|
\]
can occur; bra and ket excitation can never occur simultaneously, that is, the
simultaneous multi-transition can occur exclusively for the ket state or for
the bra state.

The time propagation factor of each mode in the state $\left(  \left|
n\right\rangle \left\langle m\right|  \right)  _{s}$ during a (positive) time
duration $t$ is given by $e^{-i(n-m)\Omega_{s}t}$ for the harmonic system
without dissipation.\ 

\begin{figure}
\begin{center}
\includegraphics[scale=0.7]{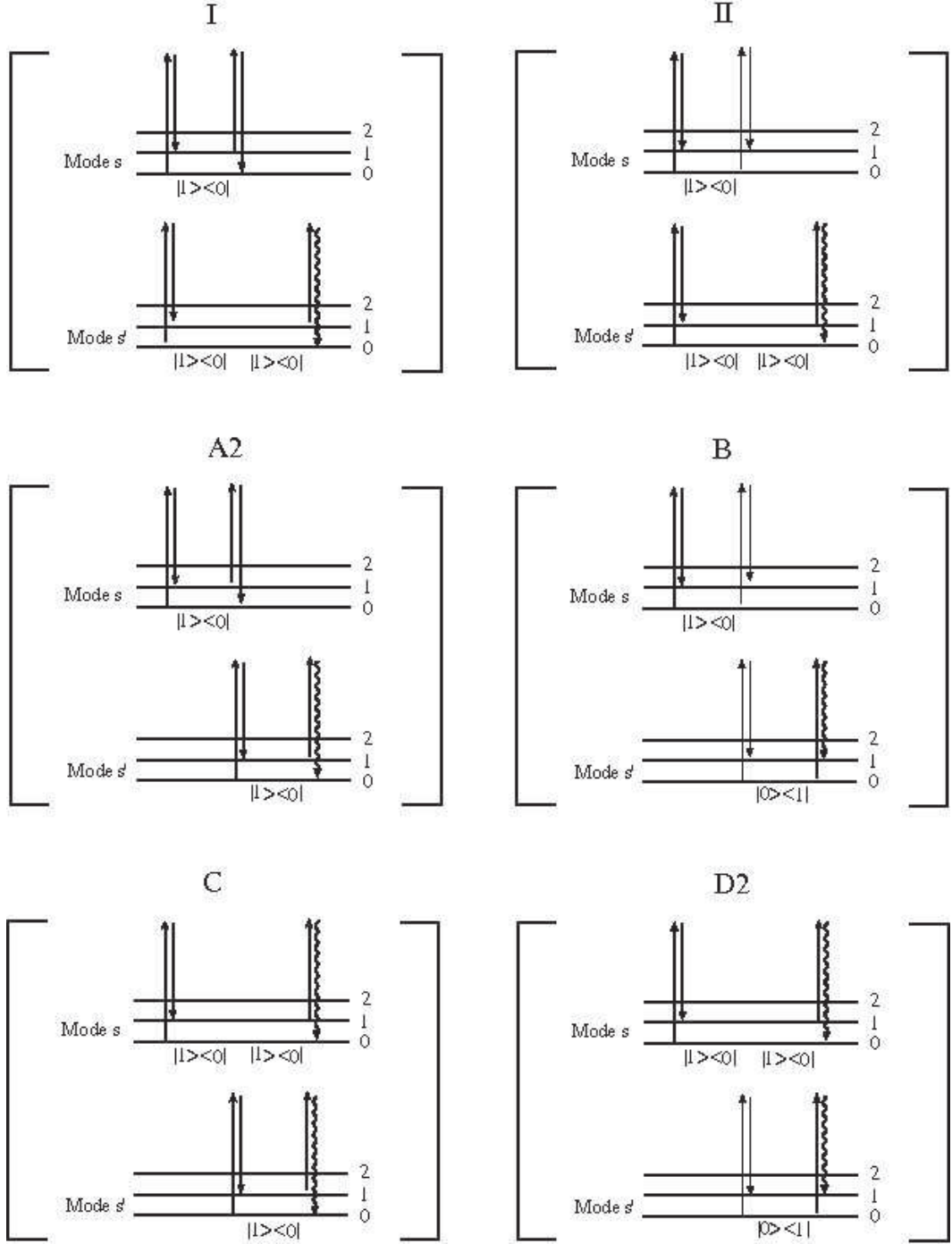}
\end{center}
\caption{Two-mode processes. It should be noted that there are no counterparts
of I', II', A1, D1.}%
\label{f8}%
\end{figure}

\begin{figure}
\begin{center}
\includegraphics[scale=0.6]{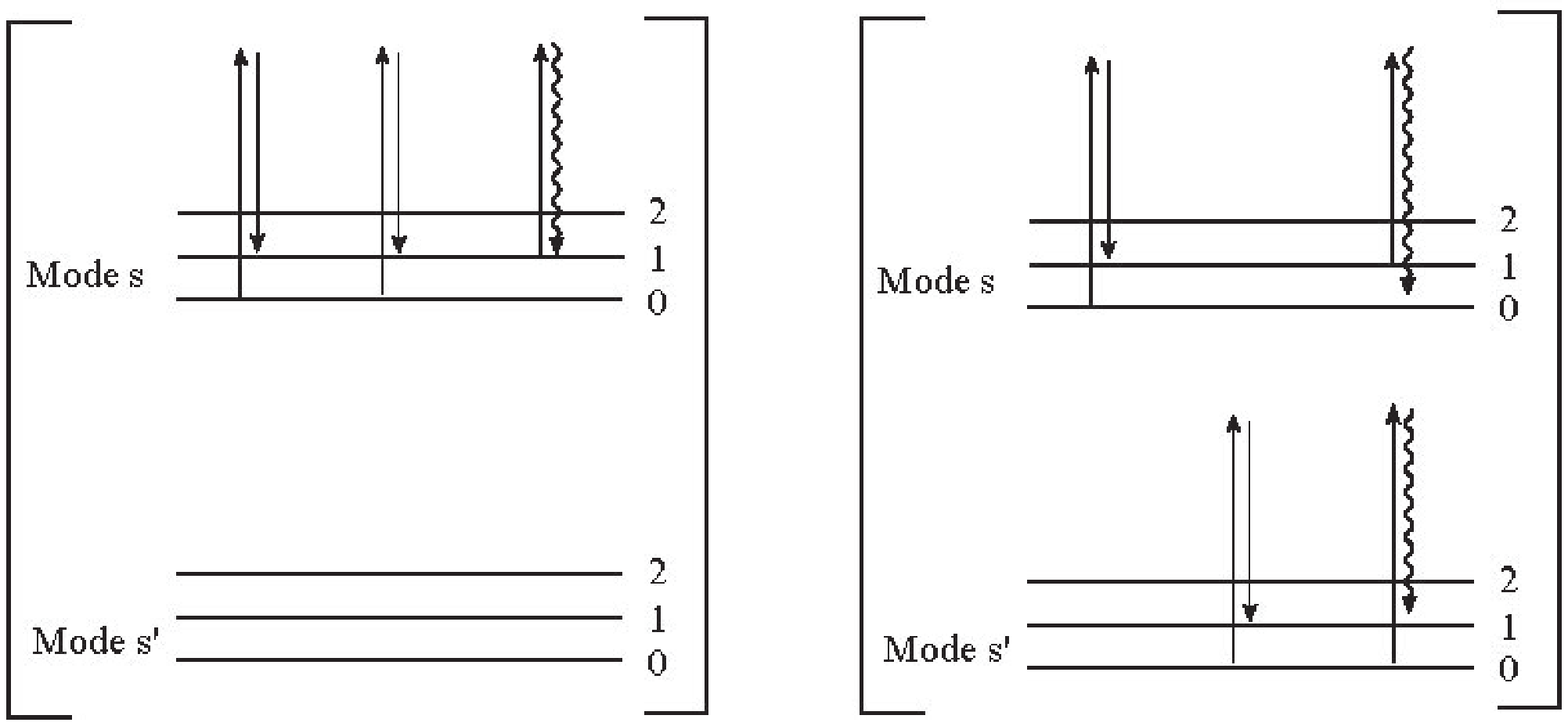}
\end{center}
\caption{Quartets represented by the quartet (D) of Fig. \ref{f7}.}%
\label{f8b}%
\end{figure}

In the multi-mode case, the diagram explicitly written in the square bracket
D2 in Fig. \ref{f7} represents either a single-mode process,%
\[
\left\{
\begin{array}
[c]{cc}%
\left\vert 1\right\rangle \left\langle 0\right\vert \rightarrow\left\vert
1\right\rangle \left\langle 1\right\vert  & \text{(mode }s\text{)}\\
\text{---}\rightarrow\text{---} & \text{(mode }s^{\prime}\text{)}%
\end{array}
\right.  ,
\]
where --- implies no time propagation, or a two-mode process,%
\begin{equation}
\left\{
\begin{array}
[c]{cc}%
\left\vert 1\right\rangle \left\langle 0\right\vert \rightarrow\left\vert
1\right\rangle \left\langle 0\right\vert  & \text{(mode }s\text{)}\\
\text{---}\rightarrow\left\vert 0\right\rangle \left\langle 1\right\vert  &
\text{(mode }s^{\prime}\text{)}%
\end{array}
\right.  , \label{D2m}%
\end{equation}
which is explicitly shown in the square braket D2 in Fig. \ref{f8}.
In other words, in the multi-mode case, quartet D2 in Fig.
\ref{f7} represents the quartets displayed in Fig. \ref{f8b}.

By use of the above rules in the multi-mode case, we see that the propagator
of the process in Eq. (\ref{D2m}) is given by $e^{-i\Omega_{s}T_{1}}\cdot
e^{-i\left(  \Omega_{s}-\Omega_{s^{\prime}}\right)  T_{2}}$because $\left\vert
1,0\right\rangle \left\langle 0,0\right\vert $ propagates for $T_{1}$ and
$\left\vert 1,0\right\rangle \left\langle 0,1\right\vert $ for $T_{2}$. The
remaining interaction factors are $\frac{i}{\hbar}x_{1}^{(s)}\left\langle
1,0\right\vert Q^{(s)}\left\vert 0,0\right\rangle \cdot\left(  -\frac{i}%
{\hbar}\right)  x_{1}^{(s^{\prime})}\left\langle 0,0\right\vert Q^{(s^{\prime
})}\left\vert 0,1\right\rangle \cdot\left(  x_{2}^{(ss^{\prime})}%
+x_{2}^{(s^{\prime}s)}\right)  \left\langle 0,1\right\vert Q^{(s)}%
Q^{(s^{\prime})}\left\vert 1,0\right\rangle /2!$ (and the factor 1/2 to avoid
double counting). Taking into account the other elements of the quartets, we
obtain the total contribution D2 of Fig. \ref{f7} in the multi-mode case in a
form:%
\begin{align}
\text{D2}  &  =4\sum_{s,s^{\prime}}\operatorname{Re}\left[  -\frac{1}%
{2}c_{ss^{\prime}}\left(  \frac{i}{\hbar}\right)  ^{2}x_{1}^{(s)}%
x_{1}^{(s^{\prime})}x_{2}^{(ss^{\prime})}\right. \label{Dm}\\
&  \left.  \frac{\hbar}{2M_{s}\Omega_{s}}\frac{\hbar}{2M_{s^{\prime}}%
\Omega_{s^{\prime}}}e^{-i\Omega_{s}T_{1}}e^{-i\left(  \Omega_{s}%
-\Omega_{s^{\prime}}\right)  T_{2}}\right] \nonumber
\end{align}
where $c_{ss^{\prime}}$ is 1 and 3/2 for $s\neq s^{\prime}$ and for
$s=s^{\prime},$ respectively. Comparing this with diagrams we learn that we
should associate $\left\vert n_{s},n_{s^{\prime}}\right\rangle \left\langle
m_{s},m_{s^{\prime}}\right\vert $ with $\Omega_{n_{s}m_{s}}^{(s)}%
+\Omega_{n_{s^{\prime}}m_{s^{\prime}}}^{(s^{\prime})}$. These 4 quartets
correspond to 4 diagrams in Fig. \ref{f8b} (in the dissipation-less case).

In this way (taking into account the effect of dissipation), we have%

\begin{align}
&  R^{(2)}\left(  T_{1},T_{2}\right) \label{R2m}\\
&  =\sum_{s=1,2}\left(  \text{I}_{s}+\text{II}_{s}+\text{B}_{s}+\text{C}%
_{s}+\text{D1}_{s}+\text{D2}_{2}\right) \nonumber\\
&  +\sum_{s,s^{\prime}}\text{A2}_{ss^{\prime}}+\left.  \sum_{s,s^{\prime}%
}\right.  ^{\prime}\left(  \text{B}_{ss^{\prime}}+\text{C}_{ss^{\prime}%
}+\text{D}_{ss^{\prime}}\right)  ,\nonumber
\end{align}
where the prime in the expression, $\sum_{s,s^{\prime}}^{\prime}$ , implies
that the terms with $s=s^{\prime}$ are excluded in the sum. Here, each term is
given by:
\begin{align*}
\text{I}_{s}  &  =-f_{ss}e^{-\Gamma_{20}^{(s)}T_{1}-\Gamma_{10}^{(s)}T_{2}%
}\cos\left(  2\zeta_{s}T_{1}+\zeta_{s}T_{2}\right) \\
\text{II}_{s}  &  =f_{ss}e^{-\Gamma_{20}^{(s)}T_{1}-\Gamma_{21}^{(s)}T_{2}%
}\cos\left(  2\zeta_{s}T_{1}+\zeta_{s}T_{2}\right) \\
\text{A2}_{ss^{\prime}}  &  =-f_{ss^{\prime}}e^{-\Gamma_{10}^{(s)}T_{1}%
-\Gamma_{10}^{(s^{\prime})}T_{2}}\cos\left(  \zeta_{s}T_{1}+\zeta_{s^{\prime}%
}T_{2}\right) \\
\text{B}_{s}  &  =f_{ss}e^{-\Gamma_{10}^{(s)}T_{1}-\Gamma_{12}^{(s)}T_{2}}%
\cos\left(  \zeta_{s}T_{1}-\zeta_{s}T_{2}\right) \\
\text{B}_{ss^{\prime}}  &  =f_{ss^{\prime}}e^{-\Gamma_{10}^{(s)}T_{1}%
-\Gamma_{01}^{(s^{\prime})}T_{2}}\cos\left(  \zeta_{s}T_{1}-\zeta_{s^{\prime}%
}T_{2}\right) \\
\text{C}_{s}  &  =-f_{ss}e^{-\Gamma_{10}^{(s)}T_{1}-\Gamma_{20}^{(s)}T_{2}%
}\cos\left(  \zeta_{s}T_{1}+2\zeta_{s}T_{2}\right) \\
\text{C}_{ss^{\prime}}  &  =-f_{ss^{\prime}}e^{-\Gamma_{10}^{(s)}T_{1}-\left(
\Gamma_{10}^{(s)}+\Gamma_{10}^{(s^{\prime})}\right)  T_{2}}\\
&  \times\cos\left(  \zeta_{s}T_{1}+\left(  \zeta_{s}-\zeta_{s^{\prime}%
}\right)  T_{2}\right) \\
\text{D1}_{s}  &  =-\frac{1}{2}f_{ss}e^{-\Gamma_{10}^{(s)}T_{1}-\Gamma
_{00}^{(s)}T_{2}}\cos\left(  \zeta_{s}T_{1}\right) \\
\text{D2}_{s}  &  =\frac{3}{2}f_{ss}e^{-\Gamma_{10}^{(s)}T_{1}-\Gamma
_{11}^{(s)}T_{2}}\cos\left(  \zeta_{s}T_{1}\right) \\
\text{D2}_{ss^{\prime}}  &  =-f_{ss^{\prime}}e^{-\Gamma_{10}^{(s)}%
T_{1}-\left(  \Gamma_{10}^{(s)}+\Gamma_{10}^{(s^{\prime})}\right)  T_{2}}\\
&  \times\cos\left(  \zeta_{s}T_{1}+\left(  \zeta_{s}-\zeta_{s^{\prime}%
}\right)  T_{2}\right)
\end{align*}
with%
\[
f_{ss^{\prime}}=\frac{x_{1}^{(s)}x_{1}^{(s^{\prime})}x_{2}^{(ss^{\prime})}%
}{2M_{s}\zeta_{s}M_{s^{\prime}}\zeta_{s^{\prime}}}.
\]
We remark the following: (1) I$_{ss^{\prime}}$ and II$_{ss^{\prime}}$ always
cancel out while I$_{s}$ and II$_{s}$ cancel out only if%
\[
\Gamma_{10}^{(s)}=\Gamma_{21}^{(s)}.
\]
(2) The sum $\left(  \text{A1}+\text{A2}\right)  _{s}$ is just given by
setting $s^{\prime}\rightarrow s$ in A2$_{ss^{\prime}}$. (3) When we put
\begin{align}
\text{ }\Gamma_{nn}^{(s)}  &  =\gamma_{s}/2\text{ for }\left\vert
n_{s}\right\rangle \left\langle n_{s}\right\vert \nonumber\\
\Gamma_{mn}^{(s)}  &  =\left\vert n_{s}-m_{s}\right\vert \gamma_{s}/2\text{
for }\left\vert n_{s}\right\rangle \left\langle m_{s}\right\vert \text{
(}n_{s}\neq m_{s}\text{),} \label{fr}%
\end{align}
the above expression reduces to the result of the fully corrected Brownian
oscillator model. If we employ the model with the relaxation constant,%
\[
\Gamma_{mn}^{(s)}=\frac{n+m}{2}\gamma_{s}%
\]
this leads a different result; one of the feature is the survival of the
single mode terms I$_{s}$ and II$_{s}.$

\section{Feynman rule in frequency domain}

In the frequency domain, we study the quantity%
\[
\int_{0}^{\infty}d\omega_{1}\int_{0}^{\infty}d\omega_{2}\,e^{i\omega_{1}%
T_{1}+i\omega_{2}T_{2}}R^{(2)}(T_{1},T_{2})
\]
The frequency domain expression is obtained by using the above propagators in
frequency domain (or, instead, directly by Fourier transformation of Eq.
(\ref{R2m})). The general propagating factor in the multi-mode case,
$e^{-\Gamma T_{1}-i\Omega T_{1}}\cdot e^{-\Gamma^{\prime}T_{2}-i\Omega
^{\prime}T_{2}},$ is, in the frequency domain, replaced by%
\begin{equation}
\frac{i}{\omega_{1}-\Omega+i\Gamma}\cdot\frac{i}{\omega_{2}-\Omega^{\prime
}+i\Gamma^{\prime}}. \label{prow}%
\end{equation}

\section{2D signal from each Liouville-space quartet}

In this section, we present two-dimensional signals from each Liouville-space
quartet separately in the fully corrected Brownian oscillator model. In the
frequency domain, since the signal is a complex number, we show the absolute
value of the signal. In the time domain, the signal is real, which is directly shown.

\subsection{Frequency Domain}

\subsubsection{Single weakly-damped mode}

\begin{figure}
\begin{center}
\includegraphics[scale=0.45]{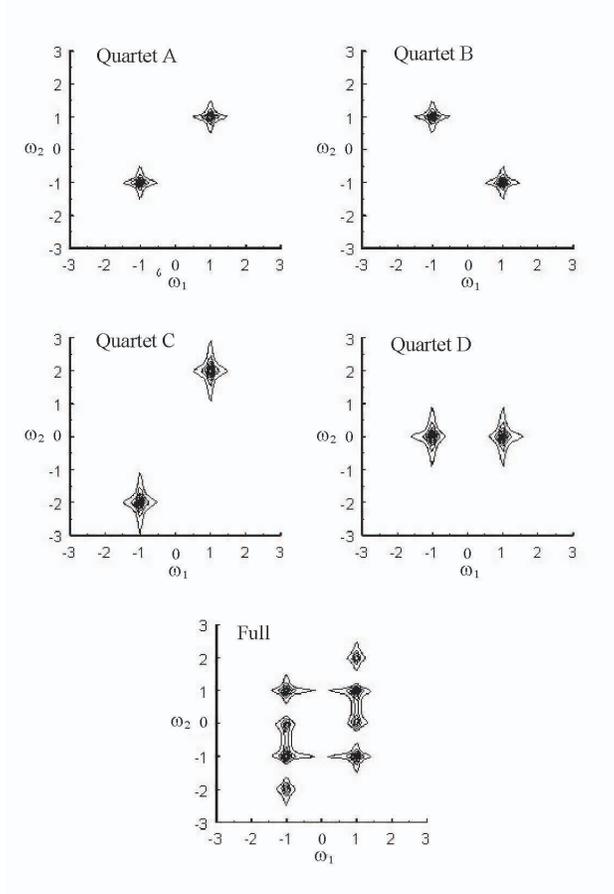}
\end{center}
\caption{Contour plot of the signal from the system with a single mode with
weak damping. The upper four plots correspond to the separate contribution
from each Liouville-space quartets. The bottom plot is the sum of them, i.e.,
the total signal.}%
\label{swd}%
\end{figure}

Fig. \ref{swd} shows signals from the system with a single mode ($\Omega=1$,
$\gamma=0.1$, in arbitrary unit). Signals from each Liouville space quartet
are separately shown. We can interpret each peak in the following way: the
process represented by $\left\vert n\right\rangle \left\langle m\right\vert
\rightarrow\left\vert n^{\prime}\right\rangle \left\langle m^{\prime
}\right\vert $ imply that the system is in the state $\left\vert
n\right\rangle \left\langle m\right\vert $ for $T_{1}$ and $\left\vert
n^{\prime}\right\rangle \left\langle m^{\prime}\right\vert $ for $T_{2}$; we
assign $\Omega_{nm}$ and $\Gamma_{nm}$ for $T_{1}$ and $\Omega_{n^{\prime
}m^{\prime}}$ and $\Gamma_{n^{\prime}m^{\prime}}$ for $T_{2}$. This can be
symbolically written as%
\[
\left\vert n\right\rangle \left\langle m\right\vert \rightarrow\left\vert
n^{\prime}\right\rangle \left\langle m^{\prime}\right\vert \Rightarrow\left\{
\begin{array}
[c]{c}%
\left(  \Omega_{nm},\Omega_{n^{\prime}m^{\prime}}\right) \\
\left(  \Gamma_{nm},\Gamma_{n^{\prime}m^{\prime}}\right)
\end{array}
\right.  .
\]
Actually, the process $\left\vert n\right\rangle \left\langle m\right\vert
\rightarrow\left\vert n^{\prime}\right\rangle \left\langle m^{\prime
}\right\vert $ corresponds to the peak at the position at $(\omega_{1}%
,\omega_{2})=\left(  \Omega_{nm},\Omega_{n^{\prime}m^{\prime}}\right)  $ with
the width in the $\omega_{1}$-axis and $\omega_{2}$-axis given by $\Gamma
_{nm}$ and $\Gamma_{n^{\prime}m^{\prime}}$, respectively. This results from
the expression in Eq. (\ref{prow}) and can be confirmed numerically as we see below.

We note here that we need not consider the contribution from the quartets I
and II because they cancels out with each other in the fully corrected
Brownian oscillator model.

Quartets A=A1+A2: this be symbolized by $\left\vert 1\right\rangle
\left\langle 0\right\vert \rightarrow\left\vert 1\right\rangle \left\langle
0\right\vert $ and its complex conjugate $\left\vert 0\right\rangle
\left\langle 1\right\vert \rightarrow\left\vert 0\right\rangle \left\langle
1\right\vert $. The former process can be symbolically written as%
\[
\left\vert 1\right\rangle \left\langle 0\right\vert \rightarrow\left\vert
1\right\rangle \left\langle 0\right\vert \Rightarrow\left\{
\begin{array}
[c]{c}%
\left(  \Omega_{10},\Omega_{10}\right) \\
\left(  \Gamma_{10},\Gamma_{10}\right)
\end{array}
\right.  \Rightarrow\left\{
\begin{array}
[c]{c}%
\left(  \Omega,\Omega\right) \\
\left(  \gamma/2,\gamma/2\right)
\end{array}
\right.  .
\]
This suggests a diagonal peak $(\omega_{1},\omega_{2})=(\Omega,\Omega)$ whose
widths in the $\omega_{1}$-direction and $\omega_{2}$-direction are both
$\gamma/2$; this peak shows symmetric pattern with respect to the two axis,
which can be seen in the contour plot in Fig. \ref{swd}. With the complex
conjugate process $\left\vert 0\right\rangle \left\langle 1\right\vert
\rightarrow\left\vert 0\right\rangle \left\langle 1\right\vert $, we associate%
\[
\left\vert 1\right\rangle \left\langle 0\right\vert \rightarrow\left\vert
1\right\rangle \left\langle 0\right\vert \Rightarrow\left\{
\begin{array}
[c]{c}%
\left(  \Omega_{01},\Omega_{01}\right) \\
\left(  \Gamma_{01},\Gamma_{01}\right)
\end{array}
\right.  \Rightarrow\left\{
\begin{array}
[c]{c}%
-\left(  \Omega,\Omega\right) \\
\left(  \gamma/2,\gamma/2\right)
\end{array}
\right.  .
\]
Namely, the quartet pair A corresponds to two symmetric diagonal peaks at
$(\omega_{1},\omega_{2})=\pm(\Omega,\Omega)$ (see the top left plot of Fig.
\ref{swd}).

Quartet B: symbolically, the association is as follows:%
\[
\left\vert 1\right\rangle \left\langle 0\right\vert \rightarrow\left\vert
1\right\rangle \left\langle 2\right\vert \Rightarrow\left\{
\begin{array}
[c]{c}%
\left(  \Omega_{10},\Omega_{12}\right) \\
\left(  \Gamma_{10},\Gamma_{12}\right)
\end{array}
\right.  \Rightarrow\left\{
\begin{array}
[c]{c}%
\left(  \Omega,-\Omega\right) \\
\left(  \gamma/2,\gamma/2\right)
\end{array}
\right.
\]
and its complex conjugate%
\[
\left\vert 0\right\rangle \left\langle 1\right\vert \rightarrow\left\vert
2\right\rangle \left\langle 1\right\vert \Rightarrow\left\{
\begin{array}
[c]{c}%
\left(  \Omega_{01},\Omega_{21}\right) \\
\left(  \Gamma_{01},\Gamma_{21}\right)
\end{array}
\right.  \Rightarrow\left\{
\begin{array}
[c]{c}%
\left(  -\Omega,\Omega\right) \\
\left(  \gamma/2,\gamma/2\right)
\end{array}
\right.
\]
Namely, we have two symmetric diagonal peaks at $(\omega_{1},\omega_{2}%
)=\pm(\Omega,-\Omega)$ (see the top right of Fig. \ref{swd}).

Quartet C: in the similar way, from the association%
\[
\left\vert 1\right\rangle \left\langle 0\right\vert \rightarrow\left\vert
2\right\rangle \left\langle 0\right\vert \Rightarrow\left\{
\begin{array}
[c]{c}%
\left(  \Omega_{10},\Omega_{20}\right) \\
\left(  \Gamma_{10},\Gamma_{20}\right)
\end{array}
\right.  \Rightarrow\left\{
\begin{array}
[c]{c}%
\left(  \Omega,-2\Omega\right) \\
\left(  \gamma/2,\gamma\right)
\end{array}
\right.
\]
and its conjugate, we should have two significant overtone peaks at
$(\omega_{1},\omega_{2})=\pm(\Omega,2\Omega)$ whose width in the $\omega_{1}%
$-direction is one half of that in the $\omega_{2}$-direction; the peak is
elongated in the second axis as can be seen in the contour plot in Fig.
\ref{swd} (see the middle left of Fig. \ref{swd}).

Quartet D=D1+D2: from the association,%
\begin{align*}
\text{D2}  &  \text{: }\left\vert 1\right\rangle \left\langle 0\right\vert
\rightarrow\left\vert 1\right\rangle \left\langle 1\right\vert \Rightarrow
\left\{
\begin{array}
[c]{c}%
\left(  \Omega_{10},\Omega_{11}\right) \\
\left(  \Gamma_{10},\Gamma_{11}\right)
\end{array}
\right.  \Rightarrow\left\{
\begin{array}
[c]{c}%
\left(  \Omega,0\right) \\
\left(  \gamma/2,\gamma\right)
\end{array}
\right. \\
\text{D1}  &  \text{: }\left\vert 1\right\rangle \left\langle 0\right\vert
\rightarrow\left\vert 0\right\rangle \left\langle 0\right\vert \Rightarrow
\left\{
\begin{array}
[c]{c}%
\left(  \Omega_{10},\Omega_{00}\right) \\
\left(  \Gamma_{10},\Gamma_{00}\right)
\end{array}
\right.  \Rightarrow\left\{
\begin{array}
[c]{c}%
\left(  \Omega,0\right) \\
\left(  \gamma/2,\gamma\right)
\end{array}
\right.
\end{align*}
and their complex conjugate, we should have two significant elongated axial
peaks at $(\omega_{1},\omega_{2})=(\pm\Omega,0)$ (see the middle right of Fig.
\ref{swd}).

The total signal displayed at the bottom of the Fig. \ref{swd} shows 8
significant peaks; now that we completely know from which Liouville-space path
each peak originates, \emph{we can assign each peak with distinct
Liouville-space paths} by the following table.

\begin{center}%
\begin{tabular}
[c]{c||c}%
quartet & peak positions in $(\omega_{1},\omega_{2})$ plane\\\hline\hline
(A) & $\left(  \Omega,\Omega\right)  ,\left(  -\Omega,-\Omega\right)  $\\
(B) & $\left(  \Omega,-\Omega\right)  ,\left(  -\Omega,\Omega\right)  $\\
(C) & $\left(  \Omega,2\Omega\right)  ,\left(  -\Omega,-2\Omega\right)  $\\
(D) & $\left(  \Omega,0\right)  ,\left(  -\Omega,0\right)  $%
\end{tabular}

\end{center}

In Fig. \ref{swd}, we notice that peaks from quartets (C) and (D) are
elongated in the second axis. This point is also understood in the above
argument, from which we have the following table.

\begin{center}%
\begin{tabular}
[c]{c||c}%
quartet & width of peaks for $(\omega_{1},\omega_{2})$\\\hline\hline
(A),(B) & $\left(  \gamma,\gamma\right)  $\\
(C),(D) & $\left(  \gamma,2\gamma\right)  $%
\end{tabular}

\end{center}

\subsubsection{Double modes (weak damping)}

\begin{figure}
\begin{center}
\includegraphics[scale=0.45]{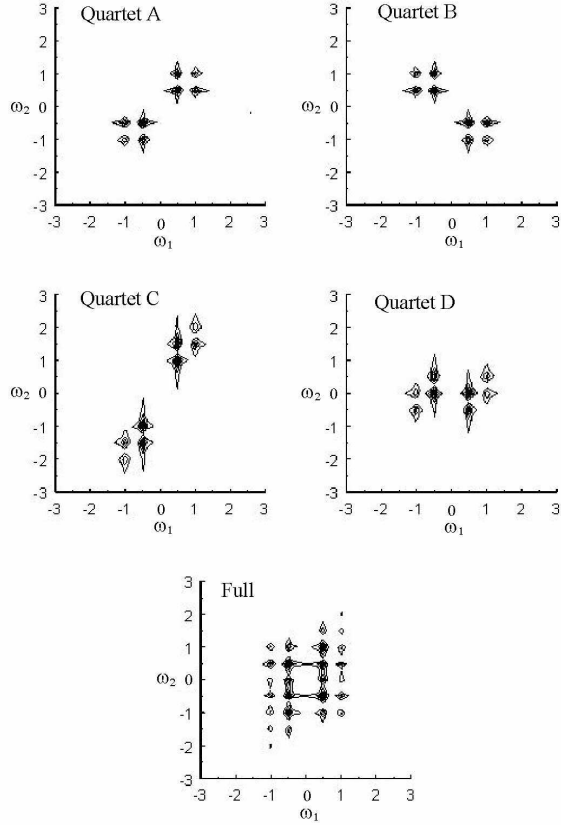}
\end{center}
\caption{Contour plot of the signal from the system with two weakly-damped
modes.}%
\label{dwd}%
\end{figure}

Fig. \ref{dwd} shows signals from the system with two weak damping modes
($\Omega_{1}=1$, $\gamma_{1}=0.1\Omega_{1}$, $\Omega_{2}=0.5$, $\gamma
_{2}=0.1\Omega_{2}$, in arbitrary unit, with the assumption, $x_{i}%
^{(s)}=M_{s}=1$, i.e., $f_{ss^{\prime}}=\left(  \zeta_{s}\zeta_{s^{\prime}%
}\right)  ^{-1}$.). Signals from each Liouville-space quartet are separately
shown. We can interpret each signal in the following way.

Top-left plot of Fig. \ref{dwd}: Two-mode quartet A2 in Fig. \ref{f8} is
associated with
\[
\left\{
\begin{array}
[c]{cc}%
\left\vert 1\right\rangle \left\langle 0\right\vert \rightarrow\text{---} &
\text{(mode }s\text{)}\\
\text{---}\rightarrow\left\vert 1\right\rangle \left\langle 0\right\vert  &
\text{(mode }s^{\prime}\text{)}%
\end{array}
\right.  \Rightarrow\left\{
\begin{array}
[c]{c}%
\left(  \Omega_{10}^{(s)},\Omega_{10}^{(s^{\prime})}\right) \\
\left(  \Gamma_{10}^{(s)},\Gamma_{10}^{(s^{\prime})}\right)
\end{array}
\right.
\]
and its complex conjugate; this quartet produces the four cross peaks at
$(\omega_{1},\omega_{2})=\pm(\Omega_{1},\Omega_{2})$, $\pm(\Omega_{2}%
,\Omega_{1})$. The remaining four diagonal peaks at $(\omega_{1},\omega
_{2})=\pm(\Omega_{1},\Omega_{1})$ and $\pm(\Omega_{2},\Omega_{2})$ originate
from the single-mode quartets A2 and A1 in Fig. \ref{f7}, which corresponds to
the process%
\[
\left\{
\begin{array}
[c]{cc}%
\left\vert 1\right\rangle \left\langle 0\right\vert \rightarrow\left\vert
1\right\rangle \left\langle 0\right\vert  & \text{(mode }s\text{)}\\
\text{---}\rightarrow\text{---} & \text{(mode }s^{\prime}\text{)}%
\end{array}
\right.
\]
and its conjugate. The widths in the $\omega_{1}$-direction and $\omega_{2}%
$-direction for the peak at $(\omega_{1},\omega_{2})=\pm(\Omega_{s}%
,\Omega_{s^{\prime}})$ are $\Gamma_{10}^{(s)}$ and $\Gamma_{10}^{(s^{\prime}%
)}$, respectively. In the fully-corrected BO model they are $\gamma_{s}/2$ and
$\gamma_{s^{\prime}}/2$, respectively. Although there exists the effect of
interferences, the relative size of the width is consistent with this
indication. For example, this is the reason the peak at (1,0.5) and (0.5, 1)
are elongated in the $\omega_{1}$ and $\omega_{2}$ axes, respectively. In
summary, in the fully-corrected BO model, the positions of peaks and
two-component of widths are given by%
\begin{align*}
\text{A2/A1 (single-mode)}  &  \text{: }\left\{
\begin{array}
[c]{c}%
\pm(\Omega_{1},\Omega_{1})\text{ with }(\gamma_{1}/2,\gamma_{1}/2)\\
\pm(\Omega_{2},\Omega_{2})\text{ with }(\gamma_{2}/2,\gamma_{2}/2)
\end{array}
\right. \\
\text{A2 (two-mode)}  &  \text{: }\left\{
\begin{array}
[c]{c}%
\pm(\Omega_{1},\Omega_{2})\text{ with }(\gamma_{1}/2,\gamma_{2}/2)\\
\pm(\Omega_{2},\Omega_{1})\text{ with }(\gamma_{2}/2,\gamma_{1}/2)
\end{array}
\right.
\end{align*}

Top-right plot of Fig. \ref{dwd}: Single-mode quartet B in Fig. \ref{f7} and
two-mode quartet B in \ref{f8} are associated with%
\begin{align*}
\left\{
\begin{array}
[c]{cc}%
\left|  1\right\rangle \left\langle 0\right|  \rightarrow\left|
1\right\rangle \left\langle 2\right|  & \text{(mode }s\text{)}\\
\text{---}\rightarrow\text{---} & \text{(mode }s^{\prime}\text{)}%
\end{array}
\right.   &  \Rightarrow\left\{
\begin{array}
[c]{c}%
\left(  \Omega_{10}^{(s)},\Omega_{12}^{(s^{\prime})}\right) \\
\left(  \Gamma_{10}^{(s)},\Gamma_{12}^{(s^{\prime})}\right)
\end{array}
\right. \\
\left\{
\begin{array}
[c]{cc}%
\left|  1\right\rangle \left\langle 0\right|  \rightarrow\text{---} &
\text{(mode }s\text{)}\\
\text{---}\rightarrow\left|  0\right\rangle \left\langle 1\right|  &
\text{(mode }s^{\prime}\text{)}%
\end{array}
\right.   &  \Rightarrow\left\{
\begin{array}
[c]{c}%
\left(  \Omega_{10}^{(s)},\Omega_{01}^{(s^{\prime})}\right) \\
\left(  \Gamma_{10}^{(s)},\Gamma_{01}^{(s^{\prime})}\right)
\end{array}
\right.
\end{align*}
The single-mode quartet produces the four diagonal peaks in the top-right
plot, while the two-mode quartet the four cross peaks. The widths in the two
directions for the diagonal peaks are given by $\left(  \Gamma_{10}%
^{(s)},\Gamma_{12}^{(s^{\prime})}\right)  $ while those for the cross peaks by
$\left(  \Gamma_{10}^{(s)},\Gamma_{12}^{(s^{\prime})}\right)  $. In summary,
we have%
\begin{align*}
\text{B (single-mode)}\text{: }  &  \left\{
\begin{array}
[c]{c}%
\pm(\Omega_{1},-\Omega_{1})\text{ with }(\gamma_{1}/2,\gamma_{1}/2)\\
\pm(\Omega_{2},-\Omega_{2})\text{ with }(\gamma_{2}/2,\gamma_{2}/2)
\end{array}
\right. \\
\text{B (two-mode)}\text{: }  &  \left\{
\begin{array}
[c]{c}%
\pm(\Omega_{1},-\Omega_{2})\text{ with }(\gamma_{1}/2,\gamma_{2}/2)\\
\pm(\Omega_{2},-\Omega_{1})\text{ with }(\gamma_{2}/2,\gamma_{1}/2)
\end{array}
\right.
\end{align*}

Middle-left plot of Fig. \ref{dwd}: Single-mode quartet C in Fig. \ref{f7} and
two-mode quartet C in \ref{f8} are associated with%
\begin{align*}
\left\{
\begin{array}
[c]{cc}%
\left\vert 1\right\rangle \left\langle 0\right\vert \rightarrow\left\vert
2\right\rangle \left\langle 0\right\vert  & \text{(mode }s\text{)}\\
\text{---}\rightarrow\text{---} & \text{(mode }s^{\prime}\text{)}%
\end{array}
\right.   &  \Rightarrow\left\{
\begin{array}
[c]{c}%
\left(  \Omega_{10}^{(s)},\Omega_{20}^{(s^{\prime})}\right) \\
\left(  \Gamma_{10}^{(s)},\Gamma_{20}^{(s^{\prime})}\right)
\end{array}
\right. \\
\left\{
\begin{array}
[c]{cc}%
\left\vert 1\right\rangle \left\langle 0\right\vert \rightarrow\left\vert
1\right\rangle \left\langle 0\right\vert  & \text{(mode }s\text{)}\\
\text{---}\rightarrow\left\vert 1\right\rangle \left\langle 0\right\vert  &
\text{(mode }s^{\prime}\text{)}%
\end{array}
\right.   &  \Rightarrow\left\{
\begin{array}
[c]{c}%
\left(  \Omega_{10}^{(s)},\Omega_{10}^{(s)}+\Omega_{10}^{(s^{\prime})}\right)
\\
\left(  \Gamma_{10}^{(s)},\Gamma_{10}^{(s)}+\Gamma_{01}^{(s^{\prime})}\right)
\end{array}
\right.
\end{align*}
The single-mode quartet produces the four overtone peaks in the middle-left
plot while the two-mode quartet the four cross peaks. In summary, we have
\begin{align*}
\text{C (single-mode)}\text{: }  &  \left\{
\begin{array}
[c]{c}%
\pm(\Omega_{1},2\Omega_{1})\text{ with }(\gamma_{1}/2,\gamma_{1})\\
\pm(\Omega_{2},2\Omega_{2})\text{ with }(\gamma_{2}/2,\gamma_{2})
\end{array}
\right. \\
\text{C (two-mode)}\text{: }  &  \left\{
\begin{array}
[c]{c}%
\pm(\Omega_{1},\Omega_{1}+\Omega_{2})\text{ with }(\gamma_{1}/2,\left(
\gamma_{1}+\gamma_{2}\right)  /2)\\
\pm(\Omega_{2},\Omega_{2}+\Omega_{1})\text{ with }(\gamma_{2}/2,\left(
\gamma_{1}+\gamma_{2}\right)  /2)
\end{array}
\right.
\end{align*}
Middle-right plot of Fig. \ref{dwd}: Single-mode quartets D1 and D2 in Fig.
\ref{f7} are associated with
\begin{align*}
\left\{
\begin{array}
[c]{cc}%
\left\vert 1\right\rangle \left\langle 0\right\vert \rightarrow\left\vert
0\right\rangle \left\langle 0\right\vert  & \text{(mode }s\text{)}\\
\text{---}\rightarrow\text{---} & \text{(mode }s^{\prime}\text{)}%
\end{array}
\right.   &  \Rightarrow\left\{
\begin{array}
[c]{c}%
\left(  \Omega_{10}^{(s)},\Omega_{00}^{(s^{\prime})}\right) \\
\left(  \Gamma_{10}^{(s)},\Gamma_{00}^{(s)}\right)
\end{array}
\right. \\
\left\{
\begin{array}
[c]{cc}%
\left\vert 1\right\rangle \left\langle 0\right\vert \rightarrow\left\vert
1\right\rangle \left\langle 1\right\vert  & \text{(mode }s\text{)}\\
\text{---}\rightarrow\text{---} & \text{(mode }s^{\prime}\text{)}%
\end{array}
\right.   &  \Rightarrow\left\{
\begin{array}
[c]{c}%
\left(  \Omega_{10}^{(s)},\Omega_{11}^{(s)}\right) \\
\left(  \Gamma_{10}^{(s)},\Gamma_{11}^{(s)}\right)
\end{array}
\right.
\end{align*}
while two-mode quartet D2 in \ref{f8} are associated with%
\[
\left\{
\begin{array}
[c]{cc}%
\left\vert 1\right\rangle \left\langle 0\right\vert \rightarrow\left\vert
1\right\rangle \left\langle 0\right\vert  & \text{(mode }s\text{)}\\
\text{---}\rightarrow\left\vert 0\right\rangle \left\langle 1\right\vert  &
\text{(mode }s^{\prime}\text{)}%
\end{array}
\right.  \Rightarrow\left\{
\begin{array}
[c]{c}%
\left(  \Omega_{10}^{(s)},\Omega_{10}^{(s)}-\Omega_{10}^{(s^{\prime})}\right)
\\
\left(  \Gamma_{10}^{(s)},\Gamma_{10}^{(s)}+\Gamma_{01}^{(s^{\prime})}\right)
\end{array}
\right.
\]
The single-mode quartet produces the four axial peaks in the middle-right plot
while the two-mode quartet the four cross peaks. In summary, we have
\begin{align*}
\text{D1/D2 (single-mode)}\text{: }  &  \left\{
\begin{array}
[c]{c}%
\pm(\Omega_{1},0)\text{ with }(\gamma_{1}/2,\gamma_{1})\\
\pm(\Omega_{2},0)\text{ with }(\gamma_{2}/2,\gamma_{2})
\end{array}
\right. \\
\text{D2 (two-mode)}\text{: }  &  \left\{
\begin{array}
[c]{c}%
\pm(\Omega_{1},\Omega_{1}-\Omega_{2})\text{ with }(\frac{\gamma_{1}}{2}%
,\frac{\gamma_{1}+\gamma_{2}}{2})\\
\pm(\Omega_{2},\Omega_{2}-\Omega_{1})\text{ with }(\frac{\gamma_{2}}{2}%
,\frac{\gamma_{1}+\gamma_{2}}{2})
\end{array}
\right.
\end{align*}
Note here in the fully-corrected BO model, we have $\Gamma_{00}^{(s)}%
=\Gamma_{11}^{(s)}$ so that the widths from the single-mode quartets D1 and D2
are the same in the above.

The total signal is displayed at the bottom of the figure; \emph{we can assign
each peak with distinct Liouville-space paths} \emph{or energy-level diagrams}
(as in Fig. \ref{f8b}).

\subsection{Time domain}

\begin{figure}
\begin{center}
\includegraphics[scale=0.45]{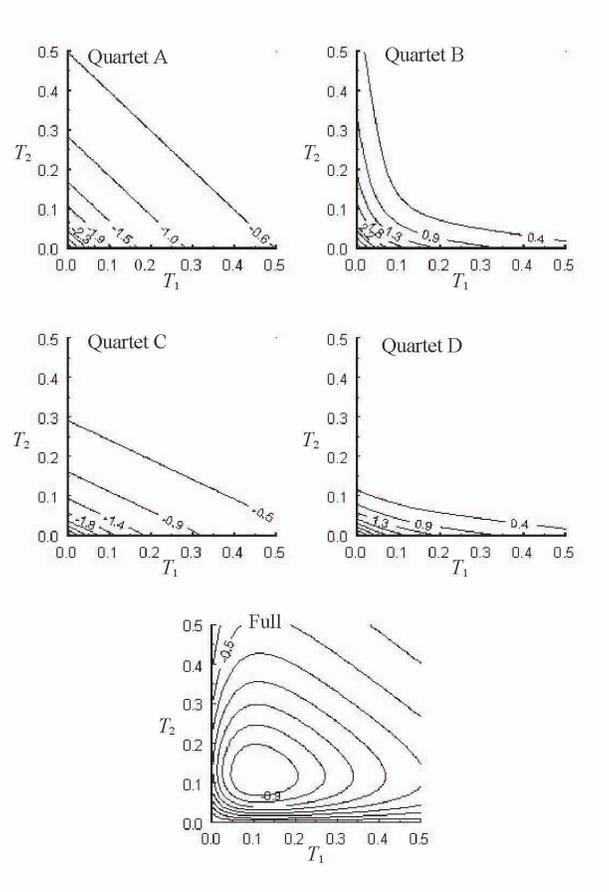}
\end{center}
\caption{Contour plot of the signal from the system with a single over-damped
mode.}%
\label{sodw}%
\end{figure}

Figs. \ref{sodw} shows the contour plots of peaks from each quartet for a
single over-damped mode system. Each quartet contributes to the total signal
rather different way. This suggests the possibility of Liouville-space-path
selective spectroscopy.

\section{Signals from Brownian oscillator model and Redfield-type model}

\begin{figure}
\begin{center}
\includegraphics[scale=0.45]{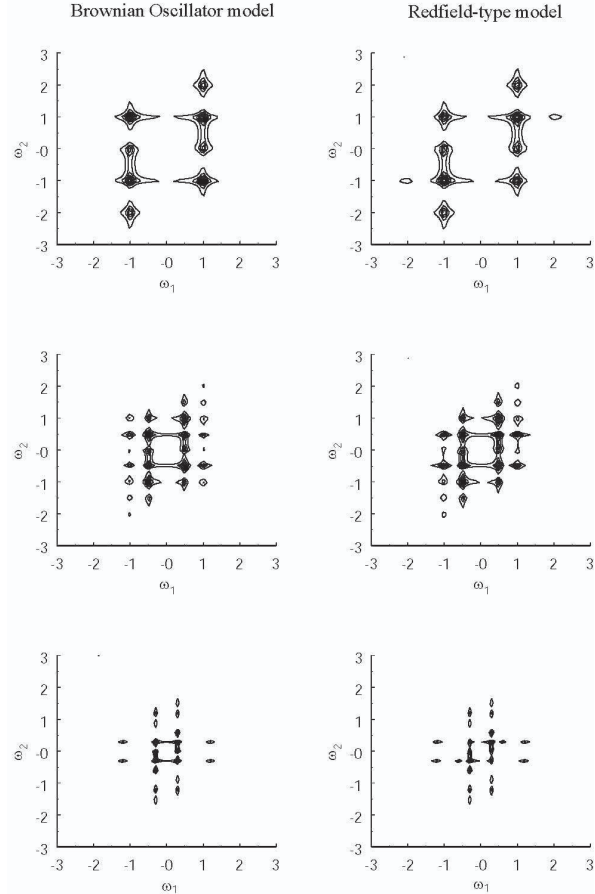}
\end{center}
\caption{2D signal from the two models. Top: single mode, $\Omega_{1}=1$.
Middle: two modes, $\Omega_{1}=1$, $\Omega_{2}=0.5$. Bottom: two modes,
$\Omega_{1}=1.2$, $\Omega_{2}=0.3$. Depending on parameters, the difference
between the models manifests as existence or non-existence of diagonal peaks.}%
\label{fbvspb}%
\end{figure}

In Fig. \ref{fbvspb}, we compare results from two models: (1) Brownian
oscillator (BO) model (the system-bath interaction is fully taken into
account) where we put Eq. (\ref{relaxB}). (2) Redfield-type model (RT) where
we put Eq. (\ref{relaxF}) with the replacement $\zeta_{s}\rightarrow\Omega
_{s}$ (no frequency shift).

Top: the right plot from RT model has extra peaks at on the left (BO) at
$(\omega_{1},\omega_{2})=\pm(2,1)$. They originate from the survival of the
quartets I and II in Fig. \ref{f7}.

Middle: on the left plot (BO) there exist extra peaks at $(\omega_{1}%
,\omega_{2})=\pm(1,-1)$. This corresponds to the single-mode quartet B in Fig.
\ref{f7}. For this process, the relaxation constants associated with the
$\omega_{2}$-axis, $\Gamma_{12}$, in BO and RT are given by $\gamma$ and
$3\gamma$, respectively; the relaxation in RT is much faster, which explains
the disappearance of the peaks. The peaks at $(\omega_{1},\omega_{2}%
)=\pm(0.5,-0.5)$ still survives because these peaks not only come from the
single-mode process B: in this case, the peaks from quartet I and II overlap
with those from other quartets.

Bottom: on the right plot (RT) exists extra peaks at $(\omega_{1},\omega
_{2})=\pm(0.3,0.6)$. They corresponds to the survival of I and II in Fig.
\ref{f7}.

In summary, the detailed situation depends on parameters. However, they have
one thing in common; the difference between the models manifests as existence
or absence of certain peaks. In the numerical results given above are all in
the weak damping regime ($\gamma\sim0.1\Omega$). The weak effect, nonetheless,
affects the existence and absence of certain peaks. This is because the
damping constants directly matter in the cancellation mechanism of certain
processes. Note that the situation is completely different for \textit{weak}
potential anharmonicity or nonlinear polarizability. Such weak effects, on the
contrary, do not concern delicate cancellation mechanisms.

If the system exhibits non-weak anharmonicity of potential or the nonlinear
system-bath coupling, as mentioned before, there may be the peaks at the
similar position predicted by the Redfield-type model. Such mechanism,
however, affect not only the existence of these peak but also the entire
profile of signal, which involves different Liouville paths. The careful study
of the signal in frequency domain shall be the critical test of the
Redfield-type model.

\section{Concluding remarks}

We stated an interpretation of the energy-level diagrams in the Liouville
space and summarize the relationships between several diagrammatic
representations. We emphasized all the diagrammatic representation reduces to
unique interpretations in Liouville space, via which we can write down
analytical expression by a Feynman rule.

We have given examples in which each Liouville process make distinctly unique
contribution to two-dimensional signal; the selective detection of quantum
process by ultrafast spectroscopy might be possible, for example, by utilizing
the phase matching condition. \cite{KTCPL} By suitably prepared spectroscopic
configuration, we might be able to concentrate on a certain quantum processes,
which allows simpler analysis and more quantitative understanding. Such
Liouville-space-path selective spectroscopy might be promising. As photon echo
can be distinguished from the pump-probe via phase matching condition, we
could differentiate spectroscopic methods by the peaks they produce.

\emph{Energy-level diagram is useful in interpreting the physical process but
it is so only after confirming the diagram certainly makes a non-zero
contribution }possibly by other method. For example, in the (fully-corrected)
Brownian oscillator model, we can assume the initial state of the diagram is
the ground state; this is because we know that other initial states result in
the same contribution from a separate calculation. Another example is the
cancellation of I by II of Fig. \ref{f7}.

In this respect, diagram in the field-theoretical context, for example,
introduced in \cite{OT} has some advantage. Number of the diagram to be
consider is considerably smaller and analytical expression is much simply
obtained; in the case of $R^{(2)}(T_{1},T_{2})$, we have only to consider just
two diagrams in total each being given by the product of two certain
propagators. This is because the cancellation is always automatically taken
into account in this method and, in addition, quartets are summed up from the
beginning in a simpler form. However, this conceals physical processes in the
Liouville space.

Note here that we have to carefully check out all possible cancellations even
if we incorporate the phase matching conditions into the response function, as
have been utilized in the electronically resonant experiments such as photon
echo and pump-probe to detect the different contributions of the Liouville
paths by choosing the laser wavevectors and frequencies. \cite{Mukamel} This
is because in vibrational spectroscopy such as IR echo, the time durations of
laser pulses are much shorter than the time periods of molecular vibrations.
\cite{KTCPL} In addition, if the initial temperature of the system is higher
than the excitation energy of vibrational levels (as in the case of low
frequency modes), or if the nonlinearity of the dipole or Raman transitions
are important, \cite{Fleming02,Saito02,OT} we have to include a number of
Liouville paths especially in higher order spectroscopy; the assignment of the
peaks to some Liouville paths become nontrivial.

As for the mechanism of relaxation, we have only considered the system
bilinearly coupled with bath. We constructed the Feynman rule by starting from
the rule in the case without damping and then by replacing the propagator so
that it causes damping with an appropriate choice of the relaxation parameters
$\Gamma_{mn}$. One may think that the set $\{\Gamma_{mn}\}$ is an arbitrary
set of parameters to fit experimental data; in the case of vibrational
spectroscopy, however, $\Gamma_{mn}$'s have to satisfy certain universal
relationships, for example, to satisfy the detailed balance condition. In
addition, the validity of the rotating wave approximation (RWA) and the
Markovian approximation associated with the second order perturbation of the
system-bath interaction might become questionable in vibrational spectroscopy;
the characterization of the relaxation processes by simple rate constants such
as $T_{1}$ and $T_{2}$ might not work. Note here that, although there are some
restrictions, one can calculate the signals without using such approximations
for Brownian model even in the anharmonic case. \cite{OTWANH,OT,T,Suzuki02} In
order to verify the consistency of theory, it is important to compare the
results from energy level models and the Brownian motion model where the
latter is based upon a microscopic picture.

In order to demonstrate how approximations for relaxation processes can change
the results, we presented the 2D signals from the Redfield-type model and
(full-order) Brownian Oscillator model, and we observed that two models give
peaks at different positions even for weak damping. This, in turn, suggests
high sensitivity to the damping mechanism of 2D spectroscopy. This situation
is in good contrast with cross peaks associated with mode-coupling of
anharmonic or nonlinear origin. They might be fairly strong to be observable
with stronger diagonal peaks. On the contrary, the cancellation mechanism is
subtle and, thus, weak damping effect can cause a drastic difference.

One of the purposes of our paper is to bridge the two complementary approaches
of the coorinate-based and the energy-level based models. The results allow us
a useful interpretation of the coordinate-based model in the energy-level
language. We should note, however, that this interpretation becomes precise
only in the weak damping limit. Nonetheless, we believe that it is useful to
have a common interpretation for the two approaches in certain situations.

\begin{acknowledgments}
We appreciate T. Kato and Y. Suzuki for a critical reading of our manuscript
prior to submission. K. O. expresses his gratitude to P.-G. de Gennes and
members of his group at Coll\`{e}ge de France, including David Qu\'{e}r\'{e},
for warm hospitality during his third stay in Paris, which is financially
supported by Coll\`{e}ge de France and Ochanomizu University. Y. T. thanks
financial support of a Grant-in-Aid for Scientific Research (B) (12440171)
from Japan Society for the Promotion of Science and Morino Science Foundation.
\end{acknowledgments}

\appendix*

\section{Expression for $R^{(2)}(\omega_{1},\omega_{2})$}

$2R^{(2)}(\omega_{1},\omega_{2})$ is given by Eq. (\ref{R2m}) with each term
expressed as%
%TCIMACRO{\TeXButton{wideb}{\begin{widetext}}}%
%BeginExpansion
\begin{widetext}%
%EndExpansion%
\begin{align*}
\text{I}_{s}  &  =\frac{f_{ss}}{\left(  \omega_{1}-z_{20}^{(s)}\right)
\left(  \omega_{2}-z_{10}^{(s)}\right)  }+\frac{f_{ss}}{\left(  \omega
_{1}+\left[  z_{20}^{(s)}\right]  ^{\ast}\right)  \left(  \omega_{2}+\left[
z_{10}^{(s)}\right]  ^{\ast}\right)  }\\
\text{II}_{s}  &  =-\frac{f_{ss}}{\left(  \omega_{1}-z_{20}^{(s)}\right)
\left(  \omega_{2}-z_{21}^{(s)}\right)  }-\frac{f_{ss}}{\left(  \omega
_{1}+\left[  z_{20}^{(s)}\right]  ^{\ast}\right)  \left(  \omega_{2}+\left[
z_{21}^{(s)}\right]  ^{\ast}\right)  }\\
&  \left(  \text{A2}\right)  _{ss^{\prime}}=\frac{f_{ss^{\prime}}}{\left(
\omega_{1}-z_{10}^{(s)}\right)  \left(  \omega_{2}-z_{10}^{(s^{\prime}%
)}\right)  }+\frac{f_{ss^{\prime}}}{\left(  \omega_{1}+\left[  z_{10}%
^{(s)}\right]  ^{\ast}\right)  \left(  \omega_{2}+\left[  z_{10}^{(s)}\right]
^{\ast}\right)  }\\
\text{B}_{s}  &  =-\frac{f_{ss}}{\left(  \omega_{1}-z_{10}^{(s)}\right)
\left(  \omega_{2}-z_{12}^{(s)}\right)  }-\frac{f_{ss}}{\left(  \omega
_{1}+\left[  z_{10}^{(s)}\right]  ^{\ast}\right)  \left(  \omega_{2}+\left[
z_{12}^{(s)}\right]  ^{\ast}\right)  }\\
\text{B}_{ss^{\prime}}  &  =-\frac{f_{ss^{\prime}}}{\left(  \omega_{1}%
-z_{10}^{(s)}\right)  \left(  \omega_{2}-z_{01}^{(s^{\prime})}\right)  }%
-\frac{f_{ss^{\prime}}}{\left(  \omega_{1}+\left[  z_{10}^{(s)}\right]
^{\ast}\right)  \left(  \omega_{2}+\left[  z_{01}^{(s)}\right]  ^{\ast
}\right)  }\\
\text{C}_{s}  &  =\frac{f_{ss}}{\left(  \omega_{1}-z_{10}^{(s)}\right)
\left(  \omega_{2}-z_{20}^{(s)}\right)  }+\frac{f_{ss}}{\left(  \omega
_{1}+\left[  z_{10}^{(s)}\right]  ^{\ast}\right)  \left(  \omega_{2}+\left[
z_{20}^{(s)}\right]  ^{\ast}\right)  }\\
\text{C}_{ss^{\prime}}  &  =\frac{f_{ss^{\prime}}}{\left(  \omega_{1}%
-z_{10}^{(s)}\right)  \left(  \omega_{2}-\left(  z_{10}^{(s)}+z_{10}%
^{(s^{\prime})}\right)  \right)  }+\frac{f_{ss^{\prime}}}{\left(  \omega
_{1}+\left[  z_{10}^{(s)}\right]  ^{\ast}\right)  \left(  \omega_{2}+\left[
z_{10}^{(s)}+z_{10}^{(s^{\prime})}\right]  ^{\ast}\right)  }\\
\left(  \text{D2}\right)  _{ss^{\prime}}  &  =-\frac{f_{ss^{\prime}}}{\left(
\omega_{1}-z_{10}^{(s)}\right)  \left(  \omega_{2}-\left(  z_{10}^{(s)}%
+z_{01}^{(s^{\prime})}\right)  \right)  }-\frac{f_{ss^{\prime}}}{\left(
\omega_{1}+\left[  z_{10}^{(s)}\right]  ^{\ast}\right)  \left(  \omega
_{2}+\left[  z_{10}^{(s)}+z_{01}^{(s^{\prime})}\right]  ^{\ast}\right)  }\\
2\cdot\text{D1}_{s}  &  =\frac{f_{ss}}{\left(  \omega_{1}-z_{10}^{(s)}\right)
\left(  \omega_{2}-z_{00}^{(s)}\right)  }+\frac{f_{ss}}{\left(  \omega
_{1}+\left[  z_{10}^{(s)}\right]  ^{\ast}\right)  \left(  \omega_{2}+\left[
z_{00}^{(s)}\right]  ^{\ast}\right)  }\\
\left(  2/3\right)  \text{D2}_{s}  &  =-\frac{2f_{ss}}{\left(  \omega
_{1}-z_{10}^{(s)}\right)  \left(  \omega_{2}-z_{11}^{(s)}\right)  }%
-\frac{2f_{ss}}{\left(  \omega_{1}+\left[  z_{10}^{(s)}\right]  ^{\ast
}\right)  \left(  \omega_{2}+\left[  z_{11}^{(s)}\right]  ^{\ast}\right)  }%
\end{align*}%
%TCIMACRO{\TeXButton{widee}{\end{widetext}}}%
%BeginExpansion
\end{widetext}%
%EndExpansion

where $z_{nm}^{(s)}=\zeta_{nm}^{(s)}-i\Gamma_{nm}^{(s)}.$

\end{document}